\documentclass[journal]{IEEEtran}
\usepackage{lineno,hyperref}
\usepackage{amsmath,amssymb}
\usepackage{float}
\newtheorem{prop}{Proposition}
\newtheorem{lemma}{Lemma}

\usepackage{graphicx}
\usepackage{multirow}
\usepackage{tabu,bm}
\usepackage{color}
\hypersetup{hidelinks}

\usepackage{algorithm}  
\usepackage{algorithmic}

\usepackage{cite}

\ifCLASSINFOpdf
  \usepackage{graphicx}
\else
\fi
\usepackage{amsmath}
\usepackage{algorithmic}

\usepackage{array}
\begin{document}
\newcommand{\tabincell}[2]{\begin{tabular}{@{}#1@{}}#2\end{tabular}}

%
\title{Non-coherent Massive SIMO Systems in ISI Channels: Constellation Design and Performance Analysis}
%
%
%

\author{Huiqiang~Xie,\IEEEmembership{}
	    Weiyang~Xu,~\IEEEmembership{Member,~IEEE}, Wei~Xiang,~\IEEEmembership{Senior Member,~IEEE}, \\Ke Shao,~\IEEEmembership{} Shengbo~Xu\IEEEmembership{}

\thanks{This work was supported by the Program for Innovation Team Building at Colleges and Universities in Chongqing, China (Grant No. CXTDX201601006) and the Key Program of Natural Science Foundation of Chongqing under Grant CSTC2017JCYJBX0047.}	    
\thanks{H. Q. Xie, W. Y. Xu and S. B. Xu are with the College of Communication Engineering, Chongqing University, Chongqing, 400044, P. R. China (E-mails: \{huiqiangxie, weiyangxu, 20134327\}@cqu.edu.cn).}
\thanks{Wei Xiang is with the College of Science, Technology \& Engineering, James Cook University, Cairns, QLD 4870, Australia (E-mail: wei.xiang@jcu.edu.au).}
\thanks{Ke Shao is with the Nanchang Institute of Technology, Nanchang, Jiangxi, 330029, P. R. China (E-mail: shaoke@163.com).}}

%
%

\markboth{ }%
{Shell \MakeLowercase{\textit{et al.}}: Non-coherent Massive SIMO Systems in ISI Channels: Constellation Design and Performance Analysis}
%



\maketitle

\begin{abstract}
A massive single-input multiple-output (SIMO) system with a single transmit antenna and a large number of receive antennas in intersymbol interference (ISI) channels is considered. Contrast to existing energy detection (ED)-based non-coherent receiver where conventional pulse amplitude modulation (PAM) is employed, we propose a constellation design which minimizes the symbol-error rate (SER) with the knowledge of channel statistics. To make a comparison, we derive the SERs of the ED-based receiver with both the proposed constellation and PAM, namely $P_{e\_opt}$ and $P_{e\_pam}$. Specifically, asymptotic behaviors of the SER in regimes of a large number of receive antennas and high signal-to-noise ratio (SNR) are investigated. Analytical results demonstrate that the logarithms of both $P_{e\_opt}$ and $P_{e\_pam}$ decrease approximately linearly with the number of receive antennas, while $P_{e\_opt}$ degrades faster. It is also shown that the proposed design is of less cost, because compared with PAM, less antennas are required to achieve the same error rate.
\end{abstract}

\begin{IEEEkeywords}
Energy detection, intersymbol interference (ISI) channel,  massive single-input multiple-output (SIMO), constellation design, symbol-error rate (SER).
\end{IEEEkeywords}

%
\IEEEpeerreviewmaketitle

\section{Introduction}
%
%
%
%


\IEEEPARstart{M}{assive} multiple-input multiple-output (MIMO) systems, where a large number of antennas are deployed at base stations (BSs) to serve a small number of users sharing the same frequency resources, have recently received a great deal of interest due to its potential gains \cite{lu2014overview,marzetta2010noncooperative}. For example, massive MIMO is energy efficient as the transmit power scales down with the number of antennas. Meanwhile, channel vectors associated with different users are asymptotically orthogonal, thus both intra- and inter-cell interference can be eliminated with simple detection or precoding algorithms \cite{Rusek2012Scaling,rappaport2013millimeter,han2017investigation}. To reap the benefits that massive MIMO offers, channel state information (CSI) is required at BSs in coherent communications. However, massive antennas make acquiring CSI much more challenging than before. In fact, the computational complexity of channel estimation is so high that estimating all channels in a timely manner becomes infeasible. In addition, the issue of pilot contamination, attributed to reusing pilots among adjacent cells, would make the problem even worse because channel estimates obtained in a given cell will be corrupted by pilots transmitted by users in the other cells\cite{truong2013effects}. 

As a promising alternative, non-coherent systems require no knowledge of instantaneous CSI at either the transmitter or receiver\cite{witrisal2009noncoherent,urkowitz1967energy,moorfeld2009multilevel}. Compared with their coherent counterparts, non-coherent receivers enjoy benefits of low complexity, low power consumption and simple structures at the cost of a sub-optimal performance\cite{wang2011weighted}. Thus, non-coherent receivers are attractive in large-scale antenna systems. Based on the non-overlapping power-space profile without CSI, an optimal decision-feedback differential detector (DFDD) provides significant gains over conventional differential detection \cite{schenk2013noncoherent,Fischer2014Noncoherent}. However, the DFDD relies on a particular channel model that cannot be exploited in general. With a large number of antennas, energy detection (ED) finds its application in non-coherent massive single-input multiple-output (SIMO) systems \cite{manolakos2014constellation}. With a non-negative pulse amplitude modulation (PAM), the transmit symbols can be decoded by averaging the received signal power across all antennas. 
{\color{black}
Since the detection/decoding is performed based on the signal energy, the system should use non-negative signal constellations. For example, non-negative PAM constellations have been documented for two different wireless standards for Millimeterwave (mmWave) short-range communication, ECMA-387 and IEEE802.15.3c \cite{Yong201160,Baykas2011IEEE}, respectively. Regarding systems with a large number of antennas, ED was proposed for mmWave communications in \cite{andersen2009overview}.}

Inspired by the seminal work in \cite{manolakos2014constellation}, non-coherent massive SIMO systems have attracted a lot of attention from the research community \cite{manolakos2014csi,chowdhury2014design,jing2016design, martinez2014energy, manolakos2016energy, jing2016energy,chowdhury2016scaling}. The symbol-error rate (SER) of the ED-based non-coherent SIMO system is derived in \cite{manolakos2014csi}, based on which a minimum distance constellation is presented. An asymptotically optimal constellation is proposed in \cite{chowdhury2014design}, and its performance gap to the optimal design could be made small with large-scale antennas or large-size constellations. In\cite{jing2016design}, two ED-based receivers are proposed, one of which analyzes the instantaneous channel energy based on Gaussian approximations of the probability density function (PDF), while the other analyzes the average channel energy with chi-square cumulative distribution function (CDF). The authors in \cite{manolakos2016energy} propose to optimize the constellation with varying levels of uncertainty on channel statistics. Also, it is proved that the non-coherent massive SIMO system satisfies the same scaling law as its coherent counterpart \cite{chowdhury2016scaling}. However, the aforementioned studies focus on intersymbol interference (ISI) free scenarios\cite{manolakos2014csi, chowdhury2014design,manolakos2016energy, jing2016design,martinez2014energy,chowdhury2016scaling}. Although an ISI channel can be transformed into multiple flat-fading channels by using OFDM, the inherent high peak-to-average-ratio and sensitivity to the carrier frequency offset present new challenges. Different from the above, the authors in \cite{jing2016energy} consider the use of the ED-based receiver in multipath environments, where a zero-forcing (ZF) equalizer is employed to remove ISI. 

In non-coherent massive SIMO systems, optimizing the constellation design can provide significant performance enhancement over conventional PAM. Accordingly, Manolakos {\it et al.} propose an optimal constellation to improve the error performance in flat-fading channels \cite{manolakos2016energy}. Although both analytical and numerical results show the potential of constellation optimization in ISI-free communications, whether it could reduce the error rate and what the optimal design would be in multipath scenarios are not clear. Towards this end, this paper focuses on designing a constellation that is able to minimize the SER for the ED-based SIMO system with ISI. The main contributions of our study can be summarized as follows: 

\begin{itemize}
\item In the presence of multipath channels, the generic SER $P_e$ of the ED-based SIMO receiver, which relates to the constellation and decision thresholds, for the case of a finite number of receive antennas is derived;\footnote{$P_e$, $P_{e\_pam}$ and $P_{e\_opt}$ refer to the error probabilities corresponding to general non-negative constellations, PAM and the proposed constellation, respectively.}
	
\item Based upon the derived closed-form expression of $P_e$, we present a constellation design and decoding scheme with the objective of minimizing the error probability. Then, the SERs of both the proposed design and PAM, namely $P_{e\_opt}$ and $P_{e\_pam}$, are derived for comparison;
	
\item Asymptotic behaviors of $P_{e\_opt}$ and $P_{e\_pam}$ in regimes of a large number of receive antennas and high signal-to-noise ratios (SNRs) are investigated in detail. It is shown that the logarithm of $P_{e\_opt}$ can be approximated as a linearly decreasing function of the number of antennas, and decreases at a faster rate than the logarithm of $P_{e\_pam}$ when more antennas are equipped. Due to the multipath effect and a finite number of antennas, both $P_{e\_opt}$ and $P_{e\_pam}$ exhibit an irreducible error floor at high SNRs. However, $P_{e\_opt}$ converges to a rate much lower than $P_{e\_pam}$ under the same condition.
\end{itemize}

The remainder of this paper is organized as follows. The system model is presented in Section II. The derivation of SER, optimal constellation design and threshold setting are detailed in Section III. Section IV presents a thorough SER performance analysis under different scenarios. Numerical simulation results are presented to show the effectiveness of our algorithm in Section V. Finally, Section VI concludes this paper.

$Notation$: $\mathbb{C}^{n \times m}$ indicates a matrix composed of complex numbers of size $n \times m$. Bold-font variables represent matrices or vectors. For a random variable $x$, $x \sim {\cal CN}(\mu,\sigma^2)$ means it follows a complex Gaussian distribution with mean $\mu$ and covariance $\sigma^2$. ${\mathbb E}[\cdot]$, $\text{Var}[\cdot]$, $\text{Cov}[\cdot]$ and ${\left\|\cdot\right\|_2}$ denote the expectation, variance, covariance and ${\cal L}_2$ norm of the argument, respectively. $(\cdot)^H$ denotes the Hermitian transpose. $\Re\{\cdot\}$ and $\Im\{\cdot\}$ separately refer to the real and imaginary parts of a complex number. $\text{sup}(\cdot)$ is the least upper bound. Finally, $\text{erf}(\cdot)$ and $\text{erfc}(\cdot)$ are taken to indicate the Gaussian error function and complementary Gaussian error function, respectively.

\section{System Model}

Consider a SIMO network consisting of a single-antenna transmitter and a receiver with $M$ antennas \cite{jing2016energy}. The channel between the transmitter and each receive antenna is modeled as a finite impulse response (FIR) filter with $L$ taps\cite{pitarokoilis2012optimality}. We assume an independent channel realization from one block to another. The received signal at time $t$ can be represented by 
\begin{equation}\label{y(t)}
{\bf y}(t) = \sum\limits_{l = 0}^{L - 1} {\bf h}_ls(t-l) + {\bf n}(t)
\end{equation}
where 
${\bf y}(t) \in {\mathbb C}^{M \times 1}$, ${\bf n}(t) \in {\mathbb C}^{M \times 1}$ indicates a complex Gaussian noise vector with elements ${n_i}$ $\sim$ ${\cal CN}(0,\sigma _n^2)$, ${\bf h}_l \in {\mathbb C}^{M \times 1}$ refers to the channel realization of the $l$th path with $h_{l,i}$ $\sim$ ${\cal CN}(0,\sigma_{h_l}^2)$, $s(t)$ denotes the transmit symbol drawn from a certain non-negative constellation, and $M$ is the number of receive antennas. Our study considers the transmit SNR, which is defined as ${\text{SNR}} = {{{\mathbb E}[ {{{\left| s(t) \right|}^2}} ]}}/{{\sigma _n^2}} = {1}/{{\sigma _n^2}}$.

We focus on the following encoding and decoding scheme. It is assumed that both the receiver and transmitter possess no knowledge of instantaneous channel and noise, but the channel and noise statistics are available, i.e., means and variances. The non-negative transmit symbol $s(t)$ is selected from a constellation set ${\cal P} = \left\{\sqrt{p_1},\sqrt{p_2}, \dots,\sqrt{p_K}\right\}$, subject to the average power constraint
\begin{equation}
\frac{1}{K}\sum\limits_{i = 1}^K{p_i} \leqslant 1
\end{equation}
where ${p_i}$  denotes the $i$th constellation point and $K$ is the constellation size. Based on the ED principle, after the received signal having been filtered, squared and integrated, the average power across all antennas can be written as 
\begin{equation}\label{energy collection}
z(t) = \frac{\left\|{\bf{y}}(t)\right\|_2^2}{M}.
\end{equation}

\begin{figure}
\centering
\includegraphics[width=80mm]{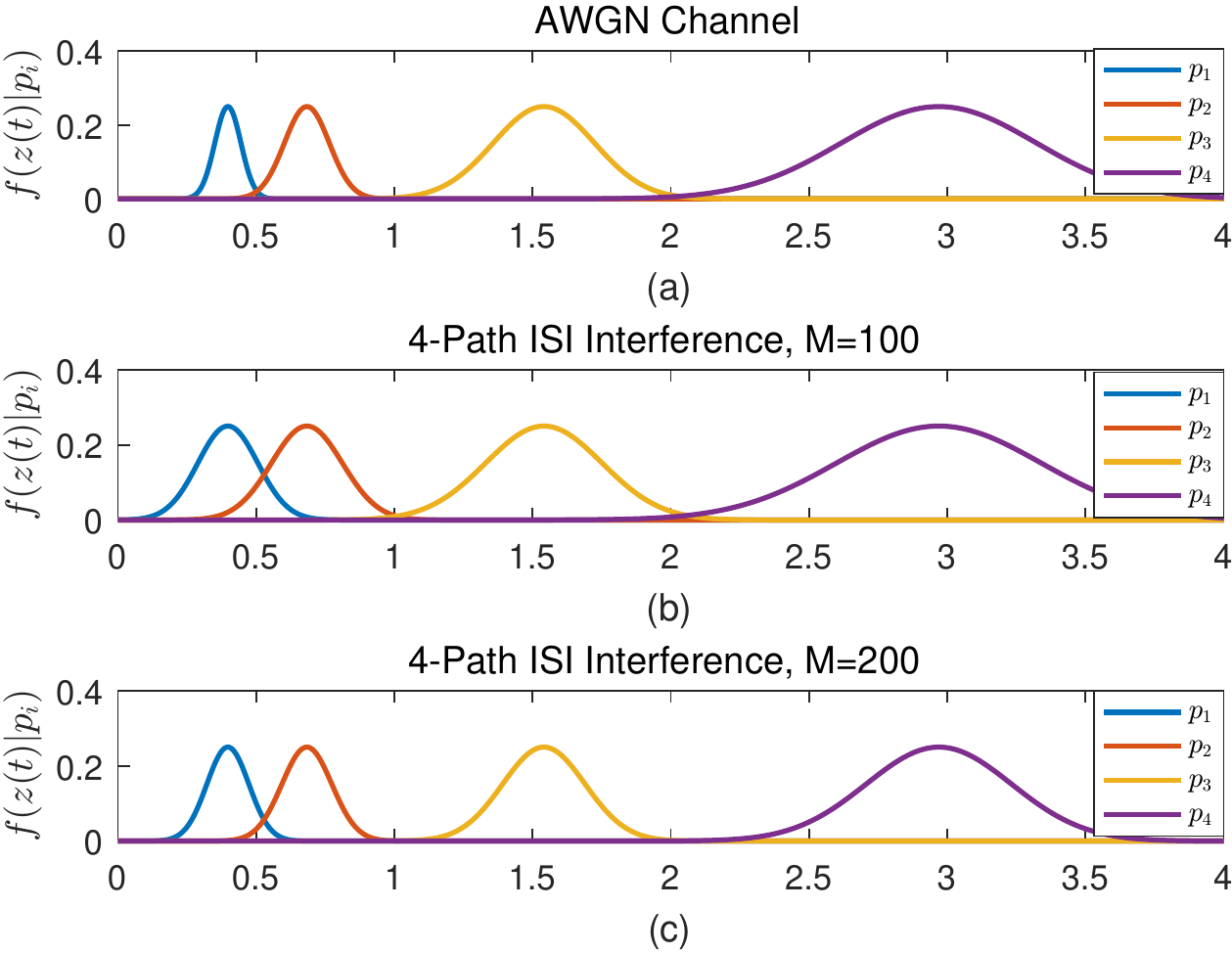}
\caption{Conditional PDFs of $z(t)$ in different scenarios, where $\rm \text{SNR}=4~dB$ for a 2-bit non-negative PAM constellation. (a) AWGN channel, $M=100$; (b) 4-path ISI channel, $M=100$; and (c) 4-path ISI channel, $M=200$.}\label{constellation distribution}\label{different_mutil_path}
\end{figure}

In flat-fading channel, $z(t)$ is taken as the decision metric for symbol detection \cite{manolakos2014constellation}. Accordingly, the positive line is partitioned into multiple decoding regions to decide which symbol was transmitted according to the observation of $z(t)$. In fact, $z(t)$ can be approximated to one of the $K$ Gaussian variables depending on $\textit{a priori}$ information of transmitted symbols. For example, with a PAM constellation of $K= \text{4}$, the PDF of $z(t)$ over an additive white Gaussian noise (AWGN) channel is shown in Fig. \ref{constellation distribution} (a), where $M=\text{100}$ and $\rm \text{SNR}= \text{4~dB}$. Clearly, four distinct Gaussian-like curves $f(z(t)|{p_i})_{i=\text{1,2,3,4}}$ can be observed, corresponding to four constellation points. As can be seen from this figure, there is a notable overlap region between $f(z(t)|{p_1})$ and $f(z(t)|{p_2})$, caused by additive noise and a finite number of antennas. Furthermore, Figs. \ref{constellation distribution} (a) and (b) indicate that the overlap region enlarges when the number of channel taps increases. This overlap will make it difficult to separate these two decoding regions, and thus decision-making between $p_1$ and $p_2$ is prone to errors. Although deploying more antennas helps reduce this overlap, as shown in Figs. \ref{constellation distribution} (b) and (c), it incurs an extra cost. Therefore, an optimal constellation is essential to reducing the error probability.

\begin{figure}
\centering
\includegraphics[width=80mm]{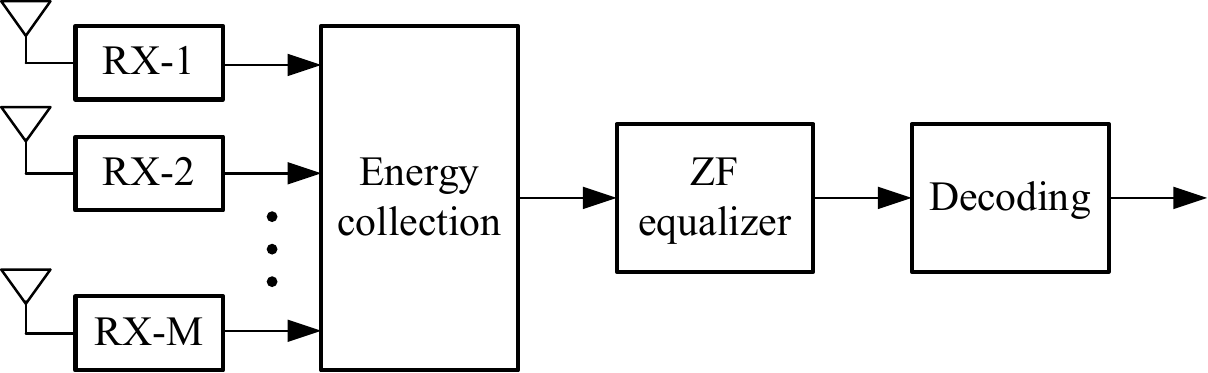}
\caption{Configuration of a non-coherent massive SIMO receiver with non-negative constellations.}
\label{system flow chart}
\end{figure}
\begin{figure}
	\centering
	\includegraphics[width=80mm]{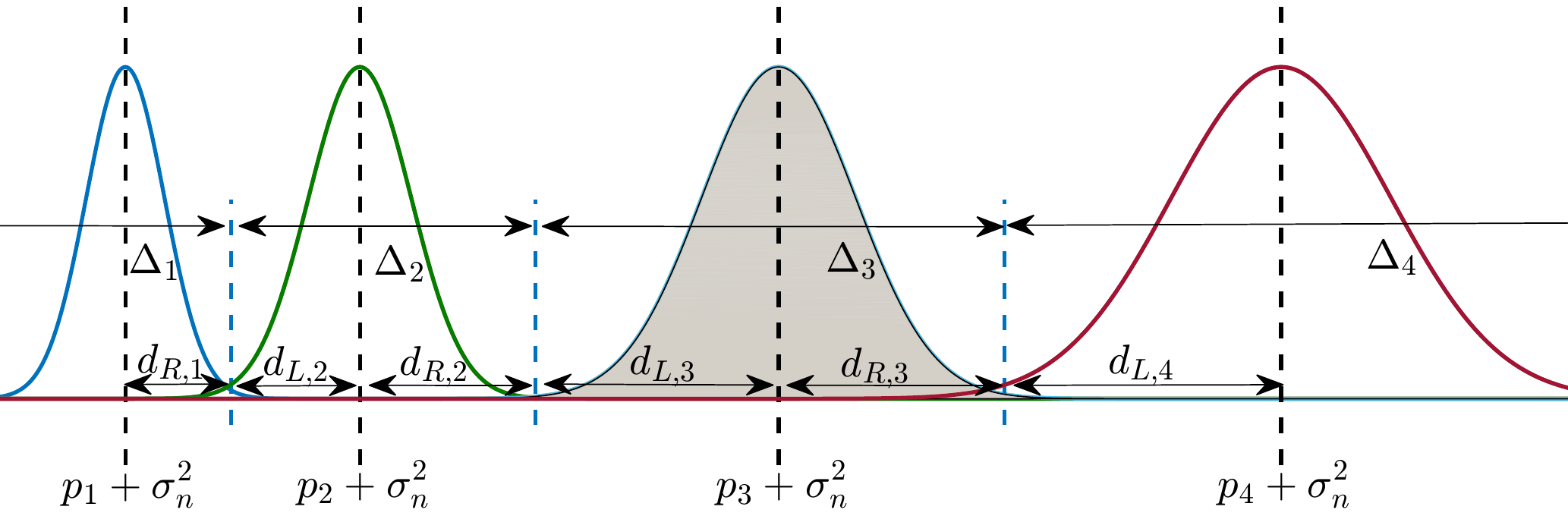}
	\caption{Decoding regions for a constellation size of $K=4$.}
	\label{constellation_picture}
\end{figure}
\section{The Proposed Constellation Design and Threshold Optimization}
In this section, a closed-form expression of SER of ED-based receivers with ISI is derived. Accordingly, our constellation design and threshold optimization is proposed to minimize the error probability.

\begin{figure*}
\begin{equation} \label{variance2}
\sigma^2_\psi ({p_i}) =  {\frac{{{w^2}}}{M}\left( {\sigma _{{h_0}}^4p_i^2 + \left( {2\sigma _{{{h_0}}}^2\sigma _{n}^2 + \frac{2}{K}\sigma _{{{h_0}}}^2\sum\limits_{l = 1}^{L - 1} {\sigma _{{h_l}}^2} } \right){p_i} + \frac{1}{{{K^2}}}\sum\limits_{l = 1}^{L - 1} {\sigma _{{h_l}}^4}  + \frac{1}{{{K^2}}}\sum\limits_{l \ne l' \ne 0} {\sigma _{{h_l}}^2} \sigma _{{h_{l'}}}^2 + \frac{2}{K}\sum\limits_{l = 1}^{L - 1} {\sigma _{{h_l}}^2\sigma _n^2 + \sigma _n^4} } \right)}
\tag{11} 
\end{equation}
\hrulefill
\end{figure*}

\subsection{SER of the ED-based Receiver in ISI Channels}
We assume that the knowledge of channel and noise statistics is available at the receiver. First, $z(t)$ in \eqref{energy collection} with a finite number of antennas can be expanded as \cite{jing2016energy}
\setcounter{equation}{3}
\begin{equation}\label{represent z(t)}
\begin{gathered}
\begin{split}
z(t) = &\frac{1}{M}\left\| {{{\mathbf{h}}_0}} \right\|_2^2{\left| {s(t)} \right|^2} + \underbrace {\frac{1}{M}\sum\limits_{l = 1}^{L - 1} {\left\| {{{\mathbf{h}}_l}} \right\|_2^2} {{\left| {s(t - l)} \right|}^2}}_{\tt IS{I_1}} \hfill \\
&+ \underbrace {\frac{1}{M}\Re \left\{ {\sum\limits_{l \ne l'} s (t - l)s(t - l'){\mathbf{h}}_l^H{{\mathbf{h}}_{l'}}} \right\}}_{\tt IS{I_2}} \hfill \\
&+ \underbrace {\frac{2}{M}\Re \left\{ {\sum\limits_{l = 0}^{L - 1} {{\mathbf{h}}_l^H{\mathbf{n}(t)}s(t - l)} } \right\}}_{\tt IS{I_3}} + \underbrace {\frac{1}{M}\left\| {{\mathbf{n}}(t)} \right\|_2^2}_{{\text{noise component}}~({\tt{NC}})}  
\end{split}
\end{gathered} 
\end{equation}
where the first component contains the desired signal. When $M \to \infty$, both $\tt ISI_2$ and $\tt ISI_3$ converge to zero and the noise component ${\tt NC}\to \sigma_n^2$, leaving $\tt ISI_1$ the only component that affects the SER. However, since $M$ can never be infinite, the non-zero $\tt ISI_2$, $\tt ISI_3$ and $\tt NC$ would adversely affect the error performance. Hence, a closed-form expression of SER, which is the basis for our constellation design, can only be accurately derived when taking into consideration of a finite number of receive antennas. The derivation of the PDF of $z(t)$, which is a non-trivial task, is required to calculate the error rate. Towards this end, we resort to the following lemma.

\begin{lemma}\label{lemma1}
If the number of receive antennas $M$ grows large, then the following approximations are attainable thanks to the Central Limit Theorem (CLT)
\begin{equation}\label{eq6}
z(t) \sim {\cal N}\left( {{\mu _z},\sigma _z^2} \right)
\end{equation}
with $\mu_z$ and variance $\sigma^2_z$ shown as follows
\begin{gather*}
\begin{aligned}
{\mu _z} &= \sigma _{{h_0}}^2{\left| {s( t )} \right|^2} + \sum\limits_{l = 1}^{L - 1} {\sigma _{{h_l}}^2} {\left| {s( {t - l} )} \right|^2} + \sigma _n^2\\
\sigma _z^2 &= \frac{1}{M}{\left( {\sum\limits_{l = 0}^{L - 1} {\sigma _{{h_l}}^2{{\left| {s( {t - l} )} \right|}^2}}  + \sigma _n^2} \right)^2}
\end{aligned}
\end{gather*}
where ${\cal N}(\mu,\sigma^2)$ indicates a real Gaussian variable.
\begin{IEEEproof}
The proof can be found in Appendix \ref{appendix A}.
\end{IEEEproof}
\end{lemma}
{\color{black}Specially, the Gaussian approximation is accurate since $M$ tends to be large, even when SNR is low. According to Lemma 1, $z(t)$ can represented as follows
\begin{equation}
	\begin{array}{l}
	\begin{aligned}
		z\left( t \right) &= {\mu _z} + {\cal N}\left( {0,\sigma _z^2} \right)\\
	&= \sigma _{{h_0}}^2{\left| {s(t)} \right|^2} + \sum\limits_{l = 1}^{L - 1} {\sigma _{{h_l}}^2} {\left| {s(t - l)} \right|^2} + \sigma _n^2 + \delta \left( t \right)
	\end{aligned}
	\end{array}
\end{equation}
where $\delta \left( t \right)$ is a nondeterministic item and $\delta \left( t \right) \sim {\cal N} \left( {0,\sigma _z^2} \right)$.}In multipath channels, a ZF equalizer is employed to remove ISI in $z(t)$ before symbol detection \cite{jing2016energy}. The ZF equalization matrix can be computed by \cite{jing2016energy}
\begin{equation}
{{\bf{w}}^{\tt ZF}} = {{\bf{e}}_d}{\left( {{{\bf{G}}^T}{\bf{G}}} \right)^{ - 1}}{{\bf{G}}^T}
\end{equation}
where ${{\bf{e}}_d}$ is an all zero vector with the $d$-th entry being unity, which how to compute $d$ can find in \cite{jing2016energy}, and 
\begin{equation*}
{\bf G} = \left[\begin{array}{ccccccc}
{\sigma _{{h_0}}^2} & {\sigma _{{h_1}}^2} & \cdots & {\sigma _{{h_{L-1}}}^2} & 0 & \cdots & 0 \\ 
0 & {\sigma _{{h_0}}^2} & {\sigma _{{h_1}}^2} & \cdots & {\sigma _{{h_{L-1}}}^2} & \cdots& 0 \\ 
\vdots  & \vdots &  \ddots  & \ddots  & \ddots  & \ddots  & \vdots \\ 
0 &  \cdots  & 0 & {\sigma _{{h_0}}^2} & {\sigma _{{h_1}}^2} & \cdots & {\sigma _{{h_{L-1}}}^2}
\end{array} \right].
\end{equation*}
where ${\bf G} \in {\mathbb{R}^{J \times \left( {J + L - 1} \right)}}$, $J$ is the length of equalizer. The analytical formula for the decision metric $\psi(t)$ is defined as
\begin{equation}\label{equation8}
\begin{array}{l}
\begin{aligned}
\psi \left( t \right) &= \sum\limits_{j = 0}^{J - 1} {w_j^{\tt ZF}z\left( {t - j} \right)} \\
&= {\left| {s\left( t \right)} \right|^2} + w\sigma _n^2 + \sum\limits_{j = 0}^{J - 1} {w_j^{\tt ZF}\delta \left( {t - j} \right)} 
\end{aligned}
\end{array}
\end{equation}
where $w_j^{\tt ZF}$ is the $j$-th elements in ${{\bf{w}}^{\tt ZF}}$. The equalizer will work well when $M \to \infty$, the third item in \eqref{equation8} will converge to zeros. However, since $M$ can never be infinite, the third item in \eqref{equation8} would adversely affect the error performance.

The configuration of a non-coherent massive SIMO receiver is shown in Fig. \ref{system flow chart}. It is worth noting that $z(t)$ after equalization, denoted by $\psi(t)$, still follows the Gaussian distribution because of the linear ZF equalizer. For ease of derivation, we denote the previous symbols $s(t - l)$ by average power $\bar s = \frac{{\sqrt {{p_1}}  +  \cdots  + \sqrt {{p_K}} }}{K}$. When the current transmit symbol $s(t)=\sqrt{p_i}$, the decision metric $\psi(t)$ follows the Gaussian distribution, i.e.
\begin{equation}
\psi ( t ) \sim {\cal N}\left( {{\mu _\psi }( {{p_i}} ),\sigma _\psi ^2( {{p_i}} )} \right)
\end{equation}
where the mean and variance of $\psi(t)$ is shown as following and in \eqref{variance2} on the top of next page
\begin{equation}
{\mu_\psi(p_i)} = {p_i} + w\sigma _n^2
\end{equation}
where $w$ is a constant computed by multiplying the equalizer coefficients with an all-one vector \cite{jing2016energy}. The aforementioned results make the derivation of SER straightforward.

\begin{prop}\label{prop1}
With a finite number of receive antennas, the SER of the ED-based receiver in ISI channels is given below
\end{prop}
\setcounter{equation}{11} 
\begin{equation}\label{SER1}
\begin{array}{l}
\begin{split}
{P_{ e}} &= 1 -\displaystyle  \frac{1}{K}\sum\limits_{i = 1}^K {P({p_i})} \\
&= 1 - \displaystyle \frac{1}{{2K}}\sum\limits_{i = 1}^K {\left({\text{erf}\left({\frac{{{d_{L,i}}}}{{\sqrt 2 \sigma_\psi ({p_i})}}} \right) + \text{erf}\left( {\frac{{{d_{R,i}}}}{{\sqrt 2 \sigma_\psi ({p_i})}}}\right)}\right)}
\end{split}
\end{array}
\end{equation}
where $d_{L,i}$ and $d_{R,i}$ are values of thresholds shown in Fig. \ref{constellation_picture}, $P(p_i)$ denotes the probability of correct decision on $p_i$. Specifically, ${d_{L,1}} = -\infty$ and ${d_{R,K}} = +\infty$. The shaded area in this figure, which is decided by $d_{L,i}$ and $d_{R,i}$, indicates the decision region for $p_i$.
\begin{IEEEproof}
The proof can be found in Appendix \ref{appendix B}.
\end{IEEEproof}

It is worth noting that the SER in \eqref{SER1} is a generalized result suitable for a variety of non-negative constellations. Given variance $\sigma_\psi({p_i})$ and decoding thresholds, one can obtain the corresponding error probability. Intuitively, some interesting remarks are made as follows:
\begin{itemize}
\item The SER over flat-fading channels can be obtained by setting $L=1$ in \eqref{SER1}. Moreover, the multipath effect of ISI channels could incur performance degradation compared with case of the flat-fading channel;   

\item Deploying more receive antennas is a straightforward and effective approach to reduce the error probability, because when $M$ grows unlimited, we have $\sigma_\psi(p_i) \to 0$, $\text{erf}\left(\frac{d_{R,i}}{\sqrt 2 \sigma_\psi(p_i)}\right)\to 1$, $\text{erf}\left(\frac{d_{L,i}}{\sqrt 2 \sigma_\psi(p_i)}\right)\to 1$, and finally $P_e \to 0$;  
	
\item If $M$ and $K$ are fixed and $\text{SNR} \to \infty$, $\sigma_\psi(p_i)$ converges to a steady state independent of $\sigma_n^2$. This is equivalent to that an error floor appears when the SNR is larger than a certain value. Therefore, a requirement of high SNRs is not critical in this scenario.
\end{itemize}

Before further analysis, the influence of a ZF equalizer on SER needs to be clarified. Although it is designed to remove the influence of ISI, the ZF equalizer causes another problem of noise enhancement. This is even worse in our study since the equalization increases the variance of ISI components in \eqref{represent z(t)} at the same time, making the decision between neighboring PDFs more prone to errors. This performance degradation can be minimized by constellation design.


\subsection{Proposed Constellation Design}
It is readily observed from \eqref{SER1} that the error probability decreases with the decrease of  $\sigma_\psi({p_i})$, which relates to $p_i$. This observation, coupled with the relationship between SER and $d_{L,i}$ or $d_{R,i}$, clearly demonstrates the potential to improve the error performance via optimizing the constellation. Obviously, minimizing the average symbol error probability equals to maximizing the probability of correct decision, i.e.

\begin{equation}
\label{maximize correct} 
\begin{aligned}
 \mathop\text{maximize}\limits_{\{{\cal P},{\vartriangle_1},\dots,{\vartriangle_K}\}}~&\frac{1}{K}\sum\limits_{i = 1}^K P(p_i),  \hfill \\
{\text{Subject to}}\quad &\frac{1}{K}\sum\limits_{i = 1}^K {{p_i}}  \leqslant 1,\;0\leqslant {p_i} \leqslant p_{i+1} \hfill \\ 
\end{aligned}
\end{equation}
where ${\vartriangle_k}$ represents the decision region $[{p_i} + w\sigma _n^2 - {d_{L,i}},{p_i} + w\sigma _n^2 + {d_{R,i}})$,
which is subject to the constraint of transmit power. However, solving \eqref{maximize correct} is not straightforward. First, the Cauchy-Schwarz inequality is utilized to simplify the problem solving process. 

\begin{lemma}\label{Cauchy Schwarz inequality}
In Euclidean space ${\displaystyle \mathbb {R} ^{n}}$ with standard inner product, the Cauchy-Schwarz inequality states that for all sequences of real numbers $a_i$ and $b_i$, we have
\begin{equation}
\displaystyle \left(\sum _{i=1}^{n}a_{i}b_{i}\right)^{2}\leqslant \left(\sum _{i=1}^{n}a_{i}^{2}\right)\left(\sum _{i=1}^{n}b_{i}^{2}\right)
\end{equation}
where the equality holds if and only if $a_{i} = k b_{i}$ for a certain constant $k \in \mathbb {R}^{+}$.
\end{lemma}

If we set $b_i=1$, the Cauchy-Schwarz inequality can be rewritten as
\begin{equation}
\frac{1}{n}\sum\limits_{i = 1}^n {{a_i} \leqslant \sqrt {\frac{{\sum\nolimits_{i = 1}^n {a_i^2} }}{n}} } 
\end{equation}
where the equality holds if and only if $a_1 = a_2=\cdots=a_n$. As a result, the maximum of \eqref{maximize correct}, or equivalently the maximum of $ \frac{1}{{2K}}\sum\nolimits_{i = 1}^K {\left({\text{erf}\left({\frac{{{d_{L,i}}}}{{\sqrt 2 \sigma _\psi({p_i})}}} \right) + \text{erf}\left( {\frac{{{d_{R,i}}}}{{\sqrt 2 \sigma_\psi ({p_i})}}}\right)}\right)}$ is achieved if $P(p_1) = P(p_2) =\cdots=P(p_K)$, which can be expanded as 
\begin{equation}\label{eq13}
\begin{array}{l}
{\text{erf}}\left( {\frac{{{d_{L,1}}}}{{\sqrt 2 {\sigma _\psi }({p_1})}}} \right) + {\text{erf}}\left( {\frac{{{d_{R,1}}}}{{\sqrt 2 {\sigma _\psi }({p_1})}}} \right) =  \hfill \\
{\text{erf}}\left( {\frac{{{d_{L,i}}}}{{\sqrt 2 {\sigma _\psi }({p_i})}}} \right) + {\text{erf}}\left( {\frac{{{d_{R,i}}}}{{\sqrt 2 {\sigma _\psi }({p_i})}}} \right) =  \hfill \\
\cdots  = {\text{erf}}\left( {\frac{{{d_{L,K}}}}{{\sqrt 2 {\sigma _\psi }({p_K})}}} \right) + {\text{erf}}\left( {\frac{{{d_{R,K}}}}{{\sqrt 2 {\sigma _\psi }({p_K})}}} \right) \hfill.
\end{array} 
\end{equation}

\begin{prop}\label{prop2}
The average symbol error probability $P_e$ is convex in the space spanned by $\cal P$.
\end{prop}
\begin{IEEEproof}
The proof can be found in \cite{Hammouda2015Performance}.
\end{IEEEproof}

According to Proposition \ref{prop2}, $P(p_i)$ is convex with respect to $p_i$ in region $(d_{L,i}$, $d_{R,i}]$. Thus, $P(p_i)$ can be maximized if
\begin{equation}\label{eq14}
{\text{erf}}\left( {\frac{{{d_{L,i}}}}{{\sqrt 2 {\sigma _\psi }({p_i})}}} \right) = {\text{erf}}\left( {\frac{{{d_{R,i}}}}{{\sqrt 2 {\sigma _\psi }({p_i})}}} \right).
\end{equation} 
Submitting \eqref{eq14} into \eqref{eq13}, the following result is obtained 
\begin{equation}\label{equation}
\begin{array}{l}
\text{erf}\left(\frac{{{d_{R,1}}}}{{\sqrt 2 \sigma_\psi\left({p_1}\right)}}\right) = \text{erf}\left(\frac{{{d_{L,2}}}}{{\sqrt 2 \sigma_\psi\left({p_2}\right)}}\right) =\\ \text{erf}\left(\frac{{{d_{R,2}}}}{{\sqrt 2 \sigma_\psi\left({p_2}\right)}}\right) = 
\dots=\text{erf}\left(\frac{{{d_{L,K}}}}{{\sqrt 2 \sigma_\psi\left({p_K}\right)}}\right).
\end{array}
\end{equation}
This equation indicates that to minimize the overall SER, the number of errors with respect to each constellation point should be the same. This observation is quite different from the case of PAM constellations. As can be observed from \eqref{equation} and Fig. \ref{constellation_picture}, two important results can be obtained, i.e.
\begin{equation}\label{important equation1}
\begin{aligned}
\frac{{{d_{R,i}}}}{{\sigma_\psi({p_i})}} &= \frac{{{d_{L,i + 1}}}}{{\sigma_\psi({p_{i + 1}})}} = \frac{{{p_{i + 1}} - {p_i}}}{{\sigma_\psi ({p_{i + 1}}) + \sigma_\psi({p_i})}} = \sqrt 2 T,\\
{d_{R,i}} &= {d_{L,i}},  \quad i=1,2,\dots,K
\end{aligned}
\end{equation}
where $T$ is defined for the ease of analysis.

Since $\text{erf}(\cdot)$ is a monotonically increasing function of its argument, maximizing $T$ is equivalent to maximizing the probability of correct decision. Thus, the optimization problem in \eqref{maximize correct} can be transformed into 
\begin{equation}\label{maximize T}
\begin{aligned}
\mathop {{\text{maximize}}}\limits_{\{ {\cal P},{\vartriangle _1}, \dots ,{\vartriangle _K}\} } \quad &T, \hfill \\
\text{Subject~to}\quad&\frac{{{p_{i + 1}} - {p_i}}}{{\sigma_\psi({p_{i + 1}}) + \sigma_\psi({p_i})}} = \sqrt 2 T \hfill ,\\
&\frac{1}{K}\sum\limits_{i = 1}^K {{p_i}}  \leqslant 1,\;0 \leqslant {p_i} \leqslant p_{i+1}. \hfill \\ 
\end{aligned} 
\end{equation}
The first constraint in \eqref{maximize T} can be rewritten as 
\begin{equation}\label{uequation}
{p_{i + 1}} - {p_i} = \sqrt 2 T( {{\sigma _\psi }( {{p_{i + 1}}} ) + {\sigma _\psi }( {{p_i}} )} ).
\end{equation}
Given a known $p_i$, ${\sigma _\psi }( {{p_i}} )$ can be calculated by \eqref{variance2}. Thus \eqref{uequation} is transformed into
\begin{equation}\label{eq19}
{\left( {{p_{i + 1}} - {p_i} - \sqrt 2 T{\sigma _\psi }( {{p_i}} )} \right)^2} = 2{T^2}\sigma _\psi ^2( {{p_{i + 1}}} ).
\end{equation}

Afterwards, with a fixed $T$ and an initial value of ${p_1}$, the problem can be converted to the following quadratic equation
\begin{equation}\label{quadratic equation}
A(T)p_{i + 1}^2 + B(T,p_i){p_{i + 1}} + C(T,p_i) = 0
\end{equation}
where
\begin{equation}
\begin{aligned}\label{ABC}
A(T) &= \frac{{{w^2}\sigma _{{h_0}}^4}}{M}-\frac{1}{{2{T^2}}}\\
B(T,p_i) &=  {\frac{{{p_i}}}{{{T^2}}} + \frac{{\sqrt 2 \sigma_\psi ({p_i})}}{T} + \frac{{2{w^2}\sigma_{{{h_0}}}^2\sigma_{n}^2}}{M}} +\frac{2w^2\sigma _{{{h_0}}}^2}{MK}\sum\limits_{l = 1}^{L - 1} {\sigma _{{h_l}}^2}\\
C(T,p_i) &= {C_1}-{C_2}(T,p_i)\\
{C_1} &= \frac{{{w^2}}}{M}\left( {\begin{array}{*{20}{l}}
\begin{gathered}
\frac{{{1}}}{{{K^2}}}\sum\limits_{l = 1}^{L - 1} {\sigma _{{h_l}}^4}  + \frac{{{1}}}{{{K^2}}}\sum\limits_{l \ne l' \ne 0} {\sigma _{{h_l}}^2} \sigma _{{h_{l'}}}^2 \hfill \\
+ \frac{{2}}{K}\sum\limits_{l = 1}^{L - 1} {\sigma _{{h_l}}^2\sigma _n^2 + \sigma _n^4}  \hfill \\ 
\end{gathered}  
\end{array}} \right)\\
{C_2}(T,p_i) &= {\left( {\frac{{{p_i}}}{{\sqrt 2 T}} + \sigma_\psi({p_i})} \right)^2}.
\end{aligned}
\end{equation}

If we set the initial value of $p_1=0$\footnote{In general, $p_1$ is initialized to be zero.}, ${p_{i + 1}}$ can be calculated iteratively. For example, \eqref{ABC} shows that both $B(T,p_i)$ and $C(T,p_i)$ relate to $\sigma_\psi ({p_i})$, thus the constraint condition $p_{i+1}$ with $T$ can be solved as follows

\begin{equation}\label{quadratic formula}
{p_{i + 1}}= \frac{{\sqrt 2 T\left( {\sqrt M {\sigma _\psi }\left( {{p_i}} \right) + w\sigma _n^2 + \displaystyle\frac{w}{K}\sum\limits_{l = 1}^{L - 1} {\sigma _{{h_l}}^2} } \right) + {p_i}}}{{\sqrt M  - \sqrt 2 Tw\sigma _{{h_0}}^2}}
\end{equation} 
 Notice that the range of $T$, which are the bounds of bisection, $0 < T < \sqrt {\frac{M}{2}} \frac{1}{{w\sigma _{{h_0}}^2}}$, guarantee the solution is real positive and are related with the number of antennas. Because the large number of antennas at the receiver, the range is enough large to find out optimize constellation solution. The result can be found in Appendix \ref{appendix C}. With the same approach, one can obtain ${p_2}, p_3,\dots, {p_K}$. Then, the optimal problem \eqref{maximize T} can be represent as 
\begin{equation}\label{maximize T1}
\begin{aligned}
\mathop {{\text{maximize}}}\limits_{\{ {\cal P},{\vartriangle _1}, \dots ,{\vartriangle _K}\} } \quad &T, \hfill \\
\text{Subject~to}\quad&\eqref{quadratic formula} \hfill ,\\
&\frac{1}{K}\sum\limits_{i = 1}^K {{p_i}}  \leqslant 1,\;0 \leqslant {p_i} \leqslant p_{i+1}. \hfill \\ 
\end{aligned} 
\end{equation}
\begin{prop}\label{prop3}
	$\sum_{i}p_{i}$ is an increasing function w.r.t. $T$.
\end{prop}
 \begin{IEEEproof}
 	The proof can be found in Appendix \ref{appendix C}.
 \end{IEEEproof}
If $\frac{1}{K}\sum_{i}p_{i}$ satisfies the power constraint, an optimal constellation is obtained. If not, $T$ needs to be adjusted accordingly.  If the average power is less than power constraint, $T$ needs to be increased. If not, $T$ will be decreased. At last, a simple method of bisection can be employed to calculate the optimal $T_{\max}$. In conclusion, the solution to \eqref{maximize T1} can be summarized in a step-by-step manner in Algorithm \ref{alg1}, where ${T_{lower}}$ and ${T_{upper}}$ indicate the lower and upper bounds of bisection range, respectively.

 \begin{algorithm}
	\caption{The solution to \eqref{maximize T}}
	\label{alg1}
	\begin{algorithmic}[1]
		\STATE Parameter initialization:\\
		$p_1=0$, $T=0$, $\frac{1}{K}\sum\nolimits_{i = 1}^K {{p_i}}  = 0$, \\
		${T_{lower}} = 0$, ${T_{upper}} = \sqrt {{M}/{2}}/({w\sigma _{h_0}^2})$;
		\WHILE {$\left| {{T_{upper}} - {T_{lower}}} \right| \geqslant 10^{-3}$ and $\left| {\frac{1}{K}\sum\nolimits_{i = 1}^K {{p_i}}  - 1} \right| \geqslant {10^{ - 3}}$}
		\STATE $T = \frac{{{T_{upper}} + {T_{lower}}}}{2}$;
		\STATE Utilizing \eqref{quadratic formula} and \eqref{ABC} to compute $p_2,p_3,\dots,p_K$;
		\IF{$\frac{1}{K}\sum\nolimits_{i = 1}^K {{p_i}}   \geqslant  1$}
		\STATE $T_{upper}=T$;
		\ELSE
		\STATE $T_{lower}=T$;
		\ENDIF
		\ENDWHILE \\
		\RETURN $p_1,p_2,\dots,p_K$ and $T$.
	\end{algorithmic}
\end{algorithm}

\begin{figure}
\centering
\includegraphics[width=85mm]{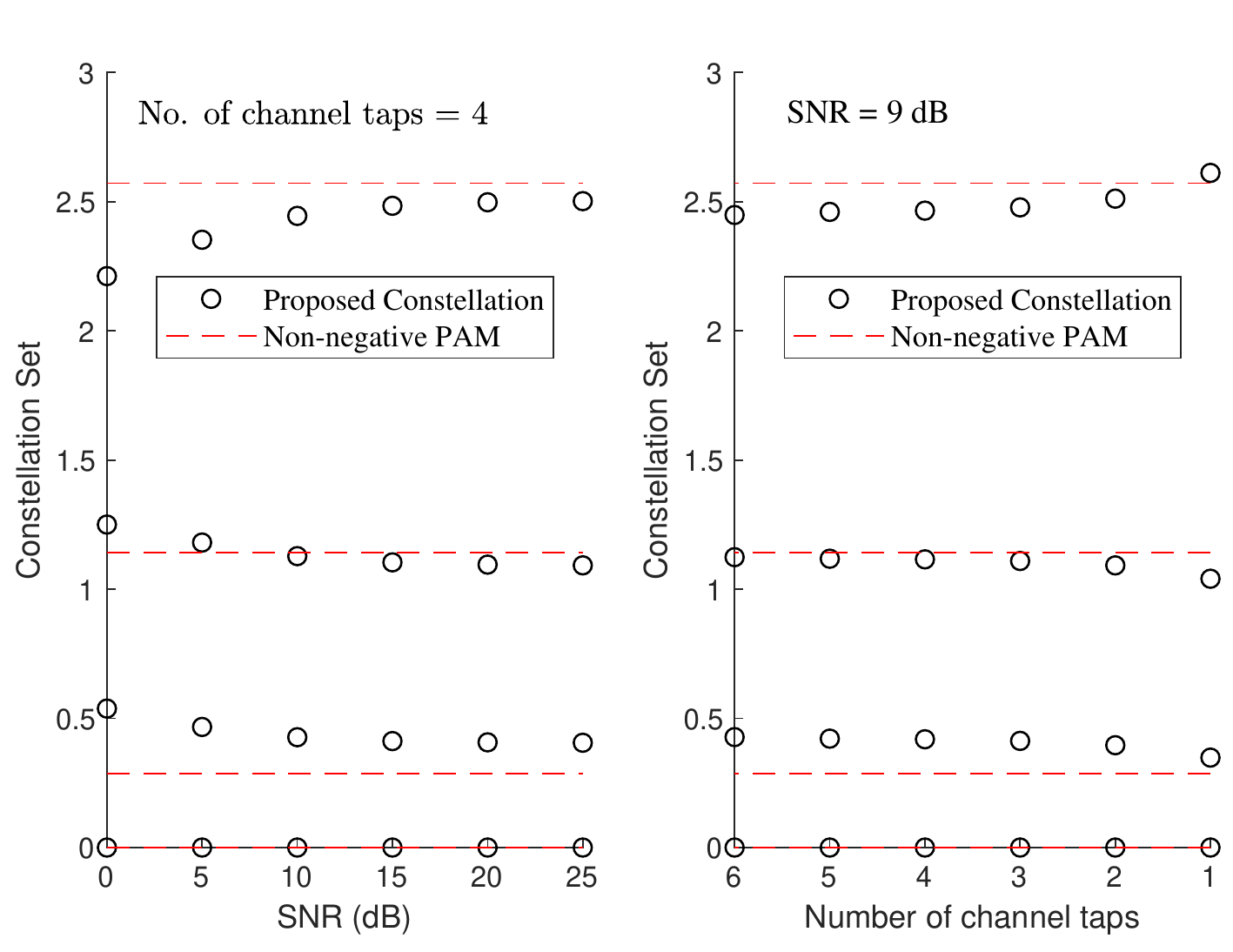}
\caption{Proposed constellation with $M=\text{100}$ and $K=\text{4}$. (a) Constellation versus the SNR; and (b) Constellation versus the number of channel taps. The red dashed line indicates the energy level of non-negative PAM.}\label{Constellation_point}
\end{figure}

Fig. \ref{Constellation_point} (a) compares the proposed constellation and a non-negative PAM at various SNRs. It is shown that the distance between $p_1$ and $p_2$ of the proposed design is larger than that of PAM at low SNRs. However, the distances between other neighboring $p_i$ of our design are smaller compared to the case of PAM. This is reasonable because the error decision between $p_1$ and $p_2$ plays the most important role in computing the SER. Moreover, our design converges to PAM gradually with the increase of SNR. Similar results is obtained if we analyze the behavior of $p_i$ when the number of channel taps varies, as shown in Fig. \ref{Constellation_point} (b). Thus, the following remarks are made:
\begin{itemize}
\item In case of any change of parameters that would result in a larger $\sigma_\psi({p_{i+1}})$, the distance between $p_1$ and $p_2$ will enlarge to make the decision between $p_1$ and $p_2$ less prone to errors;
	
\item As a result, the distances between other neighboring $p_i$ would decrease accordingly to keep the same constraint of transmit power.
\end{itemize}

In this way, the proposed design is capable of adaptively optimizing $p_i$ according to the channel and noise statistics. 



\subsection{Threshold Optimization}	
Based on \eqref{important equation1}, the decision metric can be decoded according to the maximum likelihood or other rules\cite{chowdhury2016scaling}. Given the right and left distances of ${d_{R,i}}$ and ${d_{L,i}}$ shown in Fig. \ref{constellation_picture}, the decision threshold for $s(t)={\sqrt {p_i}}$ can be obtained as
\begin{equation}
{d_{L,i}} = {d_{R,i}} = \sqrt 2 {T_{\max }}\sigma_\psi({p_i}).
\end{equation}
The optimized decision boundaries between two neighboring constellation points is then denoted as 
\begin{equation}
\begin{aligned}
tre{L_i} &= {p_i} + w\sigma _n^2 - d_{L,i}\\
tre{R_i} &= {p_i} + w\sigma _n^2 + d_{R,i}.
\end{aligned}
\end{equation}

With the optimized threshold, a transmit symbol can be decoded as follows
\begin{equation}
\hat s(t) = \left\{ \begin{gathered}
{\sqrt{p_1}},{\text{ }} \psi(t) \subseteq ( - \infty ,tre{R_1}], \hfill \\
{\sqrt{p_i}},{\text{ }} \psi(t) \subseteq (tre{L_i},tre{R_i}],{\text{ }}i = 2,3,\dots, K - 1 \hfill \\
{\sqrt{p_K}},{\text{ }} \psi(t) \subseteq (tre{L_K}, + \infty ). \hfill \\ 
\end{gathered}  \right.
\end{equation}
In the light of \eqref{equation}, the probability of correct decision is consisted of the same value
\begin{equation}
P({p_i}) = 2\text{erf}\left({T_{\max }}\right),\;i=1,2,\dots,K.
\end{equation}
Then, according to the maximized $T$ and decision thresholds, the error probability in \eqref{SER1} can be approximated as
\begin{equation}\label{SER of Optimal}
{P_{e\_opt}} \approx 1 - \frac{1}{K}\left( {\left( {K - 1} \right)\text{erf}\left({T_{\max }}\right) + 1} \right).
\end{equation}
Therefore, finding the minimum $P_{e\_opt}$ is equivalent to maximizing $T$, because $\text{erf}(\cdot)$ is a monotonically increasing function of its argument.

\subsection{Relation with the Rate Function Scheme}

Among the existing publications, the one that is most related to our proposed scheme is the one presented in \cite{manolakos2014constellation}, which is based on the rate function. This scheme was shown to be the first constellation design for non-coherent massive SIMO based on ED. However, it only explores the scenario of flat-fading channel. Motivated by this, we will next compare this scheme with the proposed design. First, we briefly review the key idea of the rate function scheme, which is represented by the following lemma.
\begin{lemma}\label{lemma rate function}
For any $d > 0$ and zero mean {\it i.i.d.} random variables $u_1, u_2,\dots,u_n$, we have\cite{manolakos2014constellation}
\begin{equation}
P\left( {\frac{{\sum\nolimits_{i = 1}^M {{u_i}} }}{M} \geqslant d} \right) \leqslant {e^{ - M\cdot I\left( d \right)}}
\end{equation}
where $I( d ) = \mathop {\sup }\limits_{\theta  > 0} \left( {\theta d - \log \left( {\mathbb{E}[e^{\theta U}]} \right)} \right)$ is the rate function.
\end{lemma}

Based on lemma \ref{lemma rate function} and moment generating function of $u_i$, the upper bound of SER can be obtained, i.e.
\begin{equation}
{P_U} \triangleq \frac{1}{K}\sum\limits_{i=1}^{K} {\left( {{e^{ - M\frac{{d_{R,i}^2}}{{2k( {{p_i}} )}}}} + {e^{ - M\frac{{d_{L,i}^2}}{{2k( {{p_i}} )}}}}} \right)}
\end{equation}
where $ k( {{p_i}} ) = {\mathbb E}[ {u_m^2} ] = \sigma _{{{\left| {{r_m}(t)} \right|}^2}}^2$, ${u_m} = {\left| {{r_m}(t)} \right|^2} - {\mu _{{{\left| {{r_m}(t)} \right|}^2}}}$ and ${r_m}(t) = h_ms(t) + n_m$ with ${r_m}(t)$ denoting the signal received by the $m$th antenna at time instant $t$, and $h_m$ being the channel between the transmitter and the $m$th antenna in the scenario of flat-fading channel.

To have a deep understanding of the difference between these two schemes, we consider the application of the proposed constellation design in flat-fading channel. Hence, the received signal and decision metric can be rewritten as
\begin{gather}
{\mathbf{r}}( t ) = {\mathbf{h}}s( t ) + {\mathbf{n}(t)}\\
{\psi_{\tt flat}}(t) = \frac{{\left\| {{\mathbf{r}}( t )} \right\|_2^2}}{M}
\end{gather}
where $\mathbf{r}(t)=[{r_1}(t),{r_2}(t),\cdots,{r_M}(t)]^H$ and $\mathbf{h} = [h_1,h_2,\dots,h_M]^H$. Based on the CLT shown in Appendix A, ${\psi_{\tt flat}}(t)$ also follows the Gaussian distribution. The relationship between $k(p_i)$ and the variance of ${\psi_{\tt flat}}(t)$, which is denoted by $\sigma_{\psi,\tt flat}^2( {{p_i}} )$, is
\begin{equation}
\sigma _{\psi ,{\tt flat}}^2({p_i}) = \frac{1}{M}\sigma _{{{\left| {{r_m}\left( t \right)} \right|}^2}}^2 = \frac{1}{M}k({p_i}).
\end{equation}

As mentioned earlier, as long as the threshold and variance are known, a closed-form expression of SER can be obtained. According to \eqref{SER1}, the SER of a non-coherent SIMO system in flat-fading channels can be written as
\begin{equation}\label{eq25}
{P_{e,\tt flat}} = \frac{1}{{2K}}\sum\limits_{i = 1}^K {\left( {{\text{erfc}}\left( {\frac{{\sqrt M {d_{L,i}}}}{{\sqrt {2k( {{p_i}} )} }}} \right) + {\text{erfc}}\left( {\frac{{\sqrt M {d_{R,i}}}}{{\sqrt {2k( {{p_i}} )} }}} \right)} \right).} 
\end{equation}

\begin{lemma}\label{lemma 3}
The complementary error function $\text{erfc}(x)$ approaches its limit when $x \to  + \infty $ as follows\cite{Oldham2008The}
\end{lemma}
\[\text{erfc}(x) \approx \frac{{e^{ - x^2}}}{{\sqrt \pi  x}},\quad {\text{if}} \quad x \to +\infty. \]

Using Lemma \ref{lemma 3}, \eqref{eq25} is able to be approximated as 
\begin{equation}\label{approximate Pe}
{P_{e,\tt flat}} \approx \frac{1}{K}\sum\limits_{i = 1}^{K} {\sqrt {\frac{{k( {{p_i}} )}}{{2\pi M}}} } \left( {\frac{{{e^{ - M\frac{{d_{R,i}^2}}{{2k( {{p_i}} )}}}}}}{{{d_{R,i}}}} + \frac{{{e^{ - M\frac{{d_{L,i}^2}}{{2k( {{p_i}} )}}}}}}{{{d_{L,i}}}}} \right).
\end{equation}
The following inequalities are easily obtained when $M$ is large
\begin{equation}\label{unequation}
\frac{{k( {{p_i}} )}}{{2\pi Md_{R,i}^2}}<1,\quad \frac{{k( {{p_i}} )}}{{2\pi Md_{L,i}^2}} < 1.
\end{equation}
As a consequence, the upper bound of $P_e$ in \eqref{approximate Pe} is
\begin{equation}\label{upper bound}
\begin{gathered}
\begin{aligned}
{P_{e,\tt flat}} <  \frac{1}{K}\sum\limits_{i=1}^{K} {\left( {{e^{ - M\frac{{d_{R,i}^2}}{{2k\left( {{p_i}} \right)}}}} + {e^{ - M\frac{{d_{L,i}^2}}{{2k\left( {{p_i}} \right)}}}}} \right)} = P_U. 
\end{aligned}
\end{gathered} 
\end{equation}
From \eqref{upper bound}, an interesting conclusion can be obtained. Through the scale of \eqref{approximate Pe}, the same upper bound of SER is observed as in \cite{manolakos2014constellation}. Therefore, the effectiveness of our proposed scheme in flat-fading channel is validated.

In conclusion, this paper provides a general framework for the constellation design in ED-based non-coherent massive SIMO systems. The results are applicable in both flat-fading and multipath-fading channels.



\section{Performance Analysis and Discussion}
In this section, we discuss the influence of key parameters on the error probability, including the number of receive antennas $M$, SNR and constellation size. The non-negative PAM and our proposed constellation design are included for comparative performance study. 

In Section III-A, we have derived the SER of the ED-based receiver with the proposed constellation design. For comparative purposes, the SER expression for a non-negative PAM constellation is given as follows
\begin{equation}\label{SER of PAM}
\begin{array}{l}
\begin{split}
{P_{e\_pam}} &= \displaystyle \frac{1}{{2K}}\sum\limits_{i = 1}^{K} {\left( {\text{erfc}\left( {\displaystyle\frac{{{d_{R\_pam,i}}}}{{\sqrt 2 \sigma_\psi ({p_{i\_pam}})}}} \right)} \right)} \\
&+ \displaystyle\frac{1}{{2K}}\sum\limits_{i = 1}^{K} {\left( {\text{erfc}\displaystyle\left( {\frac{{{d_{L\_pam,i}}}}{{\sqrt 2 \sigma_\psi({p_{i\_pam}})}}} \right)} \right)}
\end{split}
\end{array}
\end{equation}
where ${d_{R\_pam,i}} = {d_{L\_pam,i + 1}} = \frac{2i+1}{2}\varepsilon$, $\varepsilon  = \frac{6}{{\left(K - 1\right)\left(2K - 1\right)}}$. The constellation of a non-negative PAM scheme is denoted by $\sqrt {{p_{i\_pam}}}  =  \left[ {0,\sqrt \varepsilon, \dots ,i\sqrt \varepsilon, \dots, (K-1)\sqrt \varepsilon} \right]$. The result in \eqref{SER of PAM} can be obtained straightforwardly by using the same approaches as in Appendix A, with the PAM constellation and decision thresholds from \cite{jing2016energy}.

\subsection{SER Approximation}

By applying Lemma \ref{lemma 3} to \eqref{SER of Optimal} and  \eqref{SER of PAM}, it can be shown that SERs of the proposed constellation design and PAM are expressed as 
\begin{equation}\label{approximate SER of Optimal}
\begin{gathered}
\begin{aligned}
{P_{e\_opt}} &\approx \frac{{K - 1}}{K}\frac{{{e^{ - T_{\max }^2}}}}{{\sqrt \pi {T_{\max }}}}\\
{P_{e\_pam}} &\approx \frac{1}{K}\sum\limits_{i = 1}^{K} {\sqrt {\frac{1}{{2\pi }}} \frac{{\sigma_\psi ({p_{i\_pam}}){e^{ - \frac{{d_{R\_pam,i}^2}}{{2\sigma ^2_\psi({p_{i\_pam}})}}}}}}{{{d_{R\_pam,i}}}}}  \hfill \\
&\quad + \frac{1}{K}\sum\limits_{i = 1}^{K} {\sqrt {\frac{1}{{2\pi }}} \frac{{\sigma _\psi({p_{i\_pam}}){e^{ - \frac{{d_{L\_pam,i}^2}}{{2\sigma ^2_\psi({p_{i\_pam}})}}}}}}{{{d_{L\_pam,i}}}}}.  \hfill \\ 
\end{aligned}
\end{gathered} 
\end{equation}
When the constellation size of the non-negative PAM is small, e.g., $K=4$, ${P_{e\_pam}}$ is dominated by the component of $i=1$ in \eqref{approximate SER of Optimal} \cite{jing2016design}.  For example, simulation results indicate that the error decision between $p_1$ and $p_2$ accounts for up to 99.1\% of the overall errors when $M=\text{400}$ and $K=\text{4}$ at $\text{SNR}=\text{6~dB}$. Moreover, this phenomenon is also observed in Fig. \ref{different_mutil_path}. Therefore, the SER of PAM reduces to
\begin{equation}\label{L_i and R_i}
\begin{array}{l}
\begin{split}
{P_{e\_pam}} &\approx \displaystyle\frac{1}{K}\sqrt {\frac{1}{{2\pi }}} \frac{{\sigma_\psi({p_{1\_pam}}){e^{- \frac{{d_{R\_pam,1}^2}}{{2\sigma ^2_\psi({p_{1\_pam}})}}}}}}{{{d_{R\_pam,1}}}}\\
&+ \displaystyle\frac{1}{K}\sqrt {\frac{1}{{2\pi }}} \frac{{\sigma_\psi({p_{2\_pam}}){e^{ -  \frac{{d_{L\_pam,2}^2}}{{2\sigma ^2_\psi({p_{2\_pam}})}}}}}}{{{d_{L\_pam,2}}}}.
\end{split}
\end{array}
\end{equation}

It follows from \eqref{SER of PAM} that ${d_{R\_pam,i}} = {d_{L\_pam,i + 1}}$ and $\sigma_\psi({p_{2\_pam}}) > \sigma_\psi({p_{1\_pam}})$. Simulation results indicate that the error decision on $p_2$ accounts for up to 77.64\% of the overall errors when $M=\text{400}$ and $K=\text{4}$ at $\text{SNR}=\text{6~dB}$. For convenience of analysis, the first term in \eqref{L_i and R_i} is removed\footnote{This approximation in fact decreases the SER of the ED receiver employing non-negative PAM, and thus it represents the worse-case scenario for our constellation design in terms of performance comparison.}. Therefore, we can obtain the following logarithmic SER
\begin{equation}\label{log_pam}
\log {P_{e\_pam}} \approx  - \frac{{d_{L\_pam,2}^2}}{{2{\sigma ^2_\psi}({p_{2_\_pam}})}}\log e + \log \sqrt {\frac{1}{{2\pi }}} \frac{{\sigma_\psi({p_{2\_pam}})}}{{K{d_{L\_pam,2}}}}.
\end{equation}
Similarly, the logarithmic operation is applied to the SER of the proposed design, i.e.
\begin{equation}\label{logarithm optimal}
\log {P_{e\_opt}} \approx  - T_{\max }^2\log e + \log \frac{{K - 1}}{{\sqrt \pi  K{T_{\max }}}}.
\end{equation}
Based upon the results in \eqref{log_pam} and \eqref{logarithm optimal}, we now can compare the error performances between the proposed constellation design and PAM in the event of a larger number of receive antennas and high SNRs.

\subsection{Influence of a Finite Number of Receive Antennas}

\subsubsection{SER as a Function of a Finite Number of Antennas}

In order to study the influence of a finite number of receive antennas on SER, we fix the SNR and constellation size. Therefore, the variance in \eqref{variance2} can be expressed as a function of $M$
\begin{equation}\label{Variance_Antennas}
\sigma_\psi({p_i},M) = w\sqrt {\frac{1}{M}\zeta(p_i)}
\end{equation}
where $\zeta({p_i}) = \sigma_{{h_0}}^4p_i^2 + {\left( {2\sigma _{{{h_0}}}^2\sigma _{_n}^2 + \frac{2}{K}\sigma _{{{h_0}}}^2\sum\nolimits_{l = 1}^{L - 1} {\sigma _{{h_l}}^2} } \right){p_i}} + \frac{1}{{{K^2}}}\sum\nolimits_{l = 1}^{L - 1} {\sigma _{{h_l}}^4}  + \frac{1}{{{K^2}}}\sum\nolimits_{l \ne l'\ne 0} {\sigma _{{h_l}}^2} \sigma _{{h_{l'}}}^2 + \frac{2}{K}\sum\nolimits_{l = 1}^{L - 1} {\sigma _{{h_l}}^2\sigma _n^2 + \sigma _n^4}$. Since the SNR is constant in this scenario and $\sigma_{h_l}^2$ is a propagation-related parameter, $\zeta({p_i})$ only relates to ${p_i}$. It follows from \eqref{important equation1} that
\begin{equation}\label{T_max}
{T_{\max }} = \frac{{{d_{opt}}}}{{w\sqrt {2\zeta({p_{opt}})} }}\sqrt M. 
\end{equation}

Through substituting \eqref{T_max} into \eqref{logarithm optimal}, the logarithm of $P_{e\_opt}$ is expressed as a function of $M$
\begin{equation}\label{SER_optimal_antenna}
\begin{array}{l}
\begin{split}
\log {P_{e\_opt}} =&  -\displaystyle \frac{{d_{opt}^2}}{{2{w^2}\zeta({p_{opt}})}}M\log e - \frac{1}{2}\log M\\
 &+ \displaystyle\log \sqrt {\frac{2}{\pi }\zeta({p_{opt}})} \frac{{(K - 1)w}}{{K{d_{opt}}}}
\end{split}
\end{array}
\end{equation} 
where the third term is a constant irrelevant  of $M$. In addition, if $M$ grows large, the second component in \eqref{SER_optimal_antenna} plays a much less significant role compared with the first one. As a result, $\log {P_{e\_opt}}$ approximates to a linear decreasing function of $M$. This confirms that deploying more antennas is an effective way to reduce decoding errors.

\subsubsection{Non-negative PAM versus the Proposed Optimal Constellation}
Applying the same approach in \eqref{Variance_Antennas}-\eqref{SER_optimal_antenna}, the error performance of non-negative PAM can be represented as
\begin{equation}\label{SER_pam_antenna}
\begin{array}{l}
\begin{split}
\log {P_{e\_pam}} = & - \displaystyle\frac{{d_{L\_pam,2}^2}}{{2{w^2}\zeta({p_{2\_pam}})}}M\log e - \frac{1}{2}\log M\\
 &+ \displaystyle\log \sqrt {\frac{\zeta({p_{2\_pam}})}{{2\pi}}} \frac{w}{{K{d_{L\_pam,2}}}}
\end{split}
\end{array}
\end{equation}
which can also approximate to a linear decreasing function of $M$ when $M \to \infty$, the same as in \eqref{SER_optimal_antenna}. It is readily observed that the key to performance comparison between $\log {P_{e\_opt}}$ and $\log {P_{e\_pam}}$ lies in their slopes with respect to $M$. After removing constants in common, it is equivalent to comparing ${{{d_{L\_pam,2}^2}}}/{{{\zeta({p_{2\_pam}})} }}$ and ${{{d_{opt}^2}}}/{{{\zeta({p_{opt}})}}}$.

\begin{figure}
\centering
\includegraphics[width=80mm]{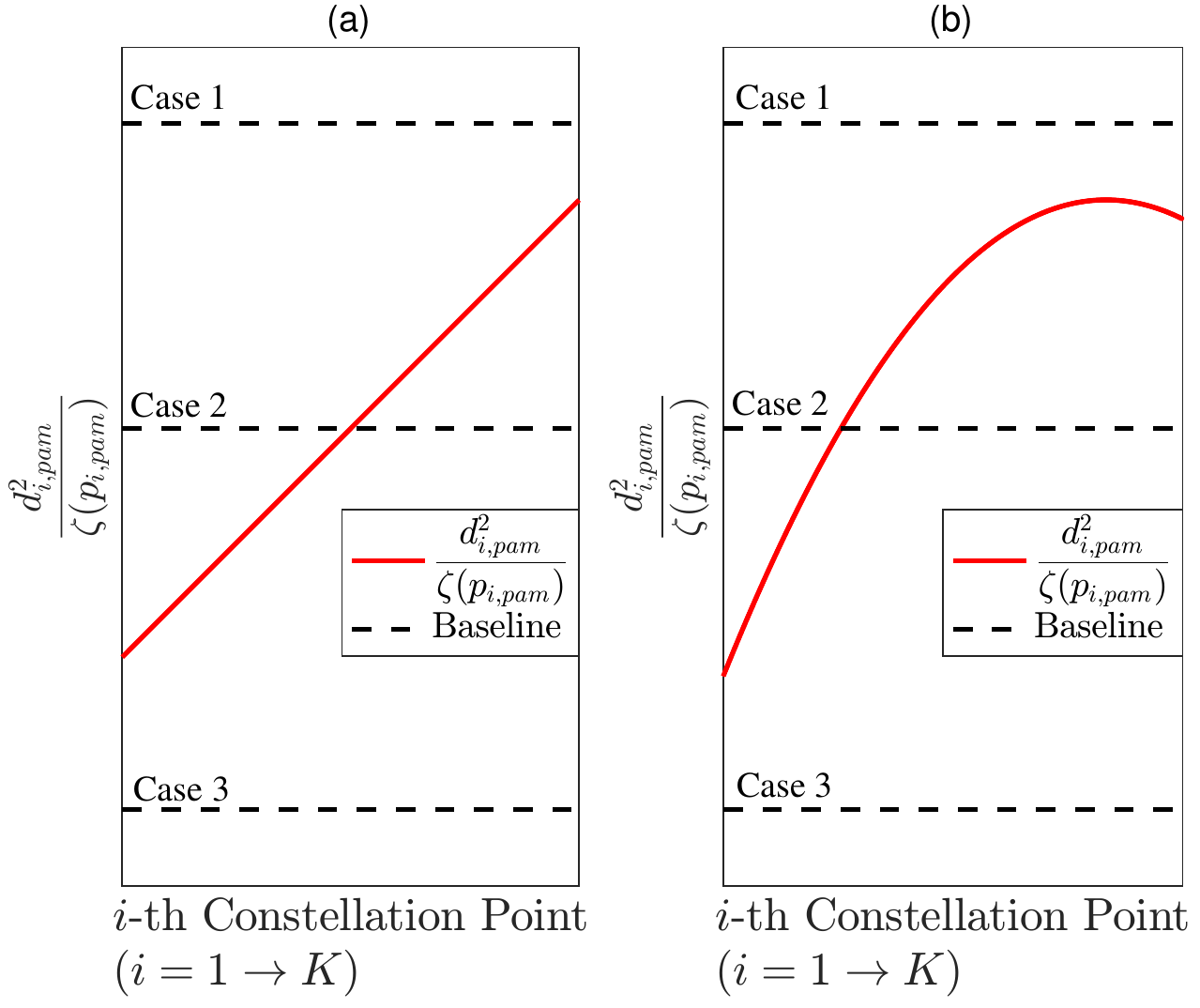}
\caption{Three situations about how the baseline crosses the curve of ${{{d_{i,pam}^2}}}/{{{\zeta({p_{i,pam}})} }}$. (a) The curve monotonically increases; and (b) The curve monotonically increases and then decreases.}\label{Baseline}
\end{figure}

First of all, it proves that ${{{d_{L\_pam,2}^2}}}/{{{\zeta({p_{2\_pam}})} }}$ is the minimum of ${{{d_{i,pam}^2}}}/{{{\zeta({p_{i\_pam}})} }}$ for non-negative PAM constellations, and monotonically increases or increases first and then decreases (refer to Appendix \ref{Appendix D} for a detailed mathematical treatment). According to the Cauchy-Schwarz inequality, ${{{d_{i,opt}^2}}}/{{{\zeta ({p_{i,opt}})}}}$ is found to be equal to ${{{d_{opt}^2}}}/{{ {\zeta({p_{opt}})}}}$ for all optimal constellation points. Accordingly, we set a baseline of ${{{d_{opt}^2}}}/{{ {\zeta({p_{opt}})}}}$ to compare with ${{{d_{L\_pam,2}^2}}}/{{{\zeta({p_{2\_pam}})} }}$. Specifically, there exist three conditions listed below, as shown in Fig. \ref{Baseline}.
\begin{itemize}
\item {\it Case 1}: If ${{{d_{L\_pam,i}^2}}}/{{{\zeta({p_{i\_pam}})}}}_{i \ne 2}$ is less than the baseline (Case 3 in Fig. \ref{Baseline} (a) and (b)), then ${{{d_{L\_pam,2}^2}}}/{{{\zeta({p_{2\_pam}})} }} <{{{d_{opt}^2}}}/{{{\zeta({p_{opt}})}}}$ because ${{{d_{L\_pam,2}^2}}}/{{{\zeta({p_{2\_pam}})} }}$ is the minimum of ${{{d_{L\_pam,i}^2}}}/{{{\zeta({p_{i\_pam}})}}}_{i=1,2,\dots,K}$;

\item {\it Case 2}: If some ${{{d_{L\_pam,i}^2}}}/{{{\zeta({p_{i\_pam}})}}}_{i \ne 2}$ are greater than the baseline (Case 2 in Fig. \ref{Baseline} (a) and (b)), we can have ${{{d_{L\_pam,2}^2}}}/{{{\zeta({p_{2\_pam}})} }} <{{{d_{opt}^2}}}/{{{\zeta({p_{opt}})}}}$ according to the power constraint;

\item {\it Case 3}: If all ${{{d_{L\_pam,i}^2}}}/{{{\zeta({p_{i\_pam}})}}}_{i=1,2,\dots,K}$ are greater than the baseline (Case 1 in Fig. \ref{Baseline} (a) and (b)), thus the SER of non-negative PAM is smaller than $P_{e\_opt}$. However, thanks to Cauchy-Schwarz inequality, $P_{e\_opt}$ is the smallest error probability. Therefore, the presumption that all ${{{d_{L\_pam,i}^2}}}/{{{\zeta({p_{i\_pam}})}}}_{i=1,2,\dots,K}$ are greater than the baseline is unfounded.   
\end{itemize}

In fact, our extensive numerical simulations reveal that only Case 2 will occur. As a consequence, $d_{L\_pam,2}^2/\zeta({p_{2\_pam}})$ are always below the baseline. 

To summarize, the result of comparing the first components of \eqref{SER_optimal_antenna} and \eqref{SER_pam_antenna} is given by
\begin{equation}\label{inequation1_antenna}
-\frac{{d_{opt}^2}}{{2{w^2}\zeta({p_{opt}})}}\log e <  - \frac{{d_{L\_pam,2}^2}}{{2{w^2}\zeta({p_{2\_pam}})}}\log e.
\end{equation}
Although the SER of the non-negative PAM maybe lower than that of the proposed design when $M$ is small, $P_{e\_opt}$ decreases at a much faster rate than $P_{e\_pam}$ with massive antennas according to \eqref{inequation1_antenna}. Hence, to maintain the same SER, our constellation design requires less antennas than PAM, which is of great benefits for practical applications.

\subsection{Influence of SNR}
\subsubsection{Non-negative PAM versus the Proposed Optimal Constellation}
In this section, the number of receive antennas and constellation size are fixed so as to investigate  the relationship between the error performance and SNR. Specifically, we will investigate the rate of descent of the SER against the SNR. In this scenario, the variance in \eqref{variance2} can be written as a function of noise variance, i.e.
\begin{equation}\label{variance_noise}
\sigma_\psi({p_i},\sigma _n^2) = \sqrt {\frac{{{w^2}}}{M}W({p_i},\sigma_n^2)}
\end{equation}
where $W({p_i},\sigma _n^2) = {\left( {\sigma _n^2 + \sigma_{{{h_0}}}^2{p_i} + \frac{1}{K}\sum\nolimits_{l = 1}^{L - 1} {\sigma _{{h_l}}^2} } \right)^2}$. Obviously, $\sigma_\psi({p_i},\sigma _n^2)$ decreases when the SNR grows. It follows from \eqref{important equation1} that 
\begin{equation}\label{SNR Tmax}
{T_{\max }} = \frac{{{d_{opt}}}}{{\sqrt 2 \sigma_\psi ({p_{opt}},\sigma _n^2)}}.
\end{equation}

Substituting \eqref{SNR Tmax} to \eqref{logarithm optimal} gives rise to
\begin{equation}\label{SNR log optimal}
\begin{gathered}
\begin{split}
\log {p_{e\_opt}} \approx&  - \frac{{d_{opt}^2}}{{2{\sigma ^2_\psi}({p_{opt}},\sigma _n^2)}}\log e\\
& - \log \frac{{{d_{opt}}}}{{\sqrt 2 \sigma_\psi ({p_{opt}},\sigma _n^2)}}+ \log \sqrt {\frac{1}{\pi }} \frac{{K - 1}}{K}.
\end{split}
\end{gathered} 
\end{equation}
Similarly, \eqref{log_pam} can be transformed into 
\begin{equation}\label{SNR log PAM}
\begin{gathered}
\begin{split}
\log {p_{e\_pam}} \approx&  - \frac{{d_{L\_pam,2}^2}}{{2{\sigma ^2_\psi}({p_{2\_pam}},\sigma _n^2)}}\log e\\
 &- \log \frac{{{d_{L\_pam,2}}}}{{\sqrt 2 \sigma_\psi ({p_{2\_pam}},\sigma _n^2)}}+ \log \sqrt {\frac{1}{\pi }} \frac{1}{K}. \hfill \\ 
\end{split}
\end{gathered}
\end{equation}
As can be observed from \eqref{SNR log optimal} and \eqref{SNR log PAM}, the first two terms follow the same model $D(x) =  - {x^2}\log e - \log x$ , while the last term is a constant independent of the SNR. Therefore, comparing the rates of descent of \eqref{SNR log optimal} and \eqref{SNR log PAM} amounts  to comparing the rates of descent of $x_{opt}={d_{opt}}/\sqrt 2 \sigma _\psi({p_{opt}},\sigma _n^2)$ and $x_{pam}={d_{L\_pam,2}}/\sqrt 2 \sigma_\psi ({p_{2\_pam}},\sigma _n^2)$ in $D(x) =  - {x^2}\log e - \log x$ with the same SNR. 

\begin{figure*}[ht]
\begin{equation}\label{variance high SNR}
\sigma^2_\psi ({p_i}) \approx  {\frac{{{w^2}}}{M}\left( 
\begin{gathered}
\sigma _{{h_0}}^4p_i^2 + \frac{2}{K}\sigma _{{{h_0}}}^2\sum\limits_{l = 1}^{L - 1} {\sigma _{{h_l}}^2} {p_i} 
+ \frac{1}{{{K^2}}}\sum\limits_{l = 1}^{L - 1} {\sigma _{{h_l}}^4}  + \frac{1}{{{K^2}}}\sum\limits_{l \ne l' \ne 0} {\sigma _{{h_l}}^2} \sigma _{{h_{l'}}}^2 \hfill \\ 
\end{gathered}  \right)}. 
\tag{58}
\end{equation}
\hrulefill
\end{figure*}

The first and second derivatives of $D(x)$ with respect to $x$ are
\begin{equation}\label{the rate of descent}
\frac{{\partial D(x)}}{{\partial x}} =  - 2x\log e - \frac{1}{{x\ln 10}}.
\end{equation}
\begin{equation}\label{a second derivative}
\frac{{{\partial ^2}D(x)}}{{\partial {x^2}}} =  - 2\log e + \frac{1}{{{x^2}\ln 10}}.
\end{equation}
Equation \eqref{the rate of descent} indicates that $D(x)$ is a monotonously decreasing function of $x$. Thanks to \eqref{a second derivative}, ${\partial D(x)}/{{\partial x}}$ monotonously increases if $0 < x < \sqrt {{1}/{{(2\log e\ln 10)}}}  \approx 0.707$. When $x > 0.707$, ${\partial D(x)}/{{\partial x}}$ monotonously decreases, which means the rate of descent at $x_1$ is larger than $x_2$ if $x_1>x_2> 0.707$. 

\begin{figure}
\centering
\includegraphics[width=80mm]{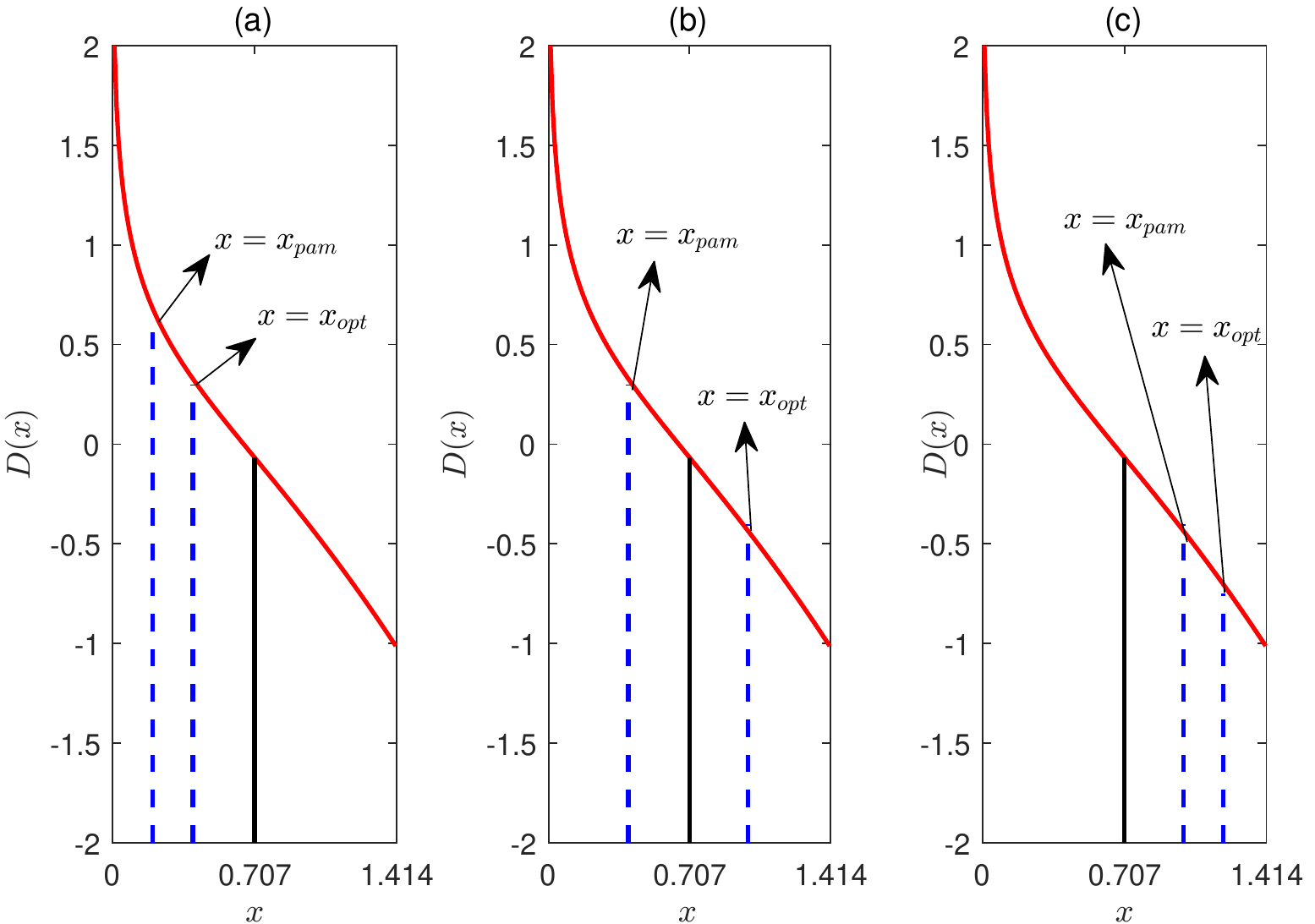}
\caption{The three situations about $x_{opt}$ and $x_{pam}$ distribution in  $D(x) =  - {x^2}\log e - \log x$. (a)  $x_{pam}<x_{opt}<0.707$, (b) $x_{pam}<0.707<x_{opt}$, (c) $0.707<x_{pam}<x_{opt}$. The red-line is $D(x) =  - {x^2}\log e - \log x$.}
\label{SNR_curve}
\end{figure}

In Section IV-B, it has already been proved that ${d_{L\_pam,2}}/\sqrt {\zeta ({p_{2\_pam}})}  < {d_{opt}}/\sqrt {\zeta ({p_{opt}})}$. The same conclusion is also applicable to this situation, i.e.
\begin{equation}\label{inequation_noise}
\frac{{{d_{L\_pam,2}}}}{{\sqrt 2 \sigma_\psi({p_{2\_pam}},\sigma _n^2)}} < \frac{{{d_{opt}}}}{{\sqrt 2 \sigma_\psi({p_{opt}},\sigma _n^2)}}.
\end{equation}

For ease of exposition, we use $x_{opt}={d_{opt}}/\sqrt 2 \sigma_\psi ({p_{opt}},\sigma _n^2)$ and $x_{pam}={d_{L\_pam,2}}/\sqrt 2 \sigma_\psi ({p_{2\_pam}},\sigma _n^2)$ in the following analysis. Specifically, there exist three cases in consideration of the distribution of $x_{opt}$ and $x_{pam}$.
\begin{itemize}
\item {\it Case 1}: $x_{pam}<x_{opt}<0.707$, shown in Fig. \ref{SNR_curve} (a). This case arises only if the number of receive antennas is very small and the SNR is low. Fig. \ref{SNR_curve} (a) demonstrates that the rate of descent of $\log {p_{e\_pam}}$ is faster than that of $\log {p_{e\_opt}}$;

\item {\it Case 2}: $x_{pam}<0.707<x_{opt}$, shown in Fig. \ref{SNR_curve} (b). The comparison between $\log {p_{e\_pam}}$ and $\log {p_{e\_opt}}$ depends on two factors, i.e., the distance to 0.707 of $x_{pam}$ and $x_{opt}$, and the rate of change of the descent of $D(x)$. Details on this case are beyond the scope of this study; and

\item {\it Case 3}: $0.707<x_{pam}<x_{opt}$, shown in Fig. \ref{SNR_curve} (c). Obviously, the descent rate of $\log {p_{e\_opt}}$ is greater than that of $\log {p_{e\_opt}}$ if $x>0.707$. With the increase of SNR, the rates of descent of both $\log {p_{e\_pam}}$ and $\log {p_{e\_opt}}$ will become even greater. 
\end{itemize}

According to our extensive simulations, Cases 1 and 2 occur when $M<50$ and $\text{SNR}<\rm -6~dB$. However, in the presence of massive receive antenna array, Case 3 would be the case. Consequently, it can be inferred that to meet the same SNR requirement, our design requires a smaller transmit power than PAM constellations.

\subsubsection{High SNR Analysis}
When the SNR becomes large enough, we have $\sigma_n^2 \to 0$ and the noise variance can be safely removed. Then, the variance in \eqref{variance2} reduces to \eqref{variance high SNR}. The second and third terms in \eqref{variance high SNR} are constants. For a fixed $M$ and constellation size, the constellation design is also fixed. Since we have already introduced a baseline, ${T_{\max }}$ can be expressed as the value of the first constellation $p_1=0$, i.e.
\setcounter{equation}{58}
\begin{equation}\label{T high SNR}
{T_{\max }} = \frac{{{d_{R,1}}K\sqrt M }}{{w\sqrt {\left( {\sum\limits_{l = 1}^{L - 1} {\sigma _{{h_l}}^4}  + \sum\limits_{l \ne l' \ne 0} {\sigma _{{h_l}}^2} \sigma _{{h_{l'}}}^2} \right)} }}
\end{equation}
where $d_{R,1}$ is the right threshold of $p_1$, as shown in Fig. \ref{constellation_picture}. 

Equation \eqref{T high SNR} shows that for a fixed $M$, ${T_{\max }}$ converges to a constant in the high SNR region. Therefore, an error floor would arise so that no matter how large the transmit power becomes, the error performance cannot be further improved according to \eqref{SER of Optimal}. There are two ways to alleviate but not eliminate the error floor. The first approach is to employ more receive antennas, which increases the numerater of ${T_{\max }}$, and then lowers the SER when the error floor appears. The second one is to alleviate the frequency-selectivity of the multipath channel, such as employing the OFDM technique. In this way, the denominator of ${T_{\max}}$ can be decreased, and the error probability converges to a steady state with lower values. Moreover, since an error floor is inevitable in this situation, high SNRs may not be required.


\section{Simulation Results}

To demonstrate the performance of the proposed constellation design, this section presents numerical results obtained via Monte Carlo simulations. The receiver structure shown in Fig. \ref{system flow chart} is applied. Non-negative PAM is considered as a benchmark \cite{jing2016design}. Throughout simulations, a 4-tap ISI channel with an exponential decay model is employed \cite{pitarokoilis2012optimality}. For both scenarios with the proposed constellation and PAM, the energy is first collected by the massive antenna array and then a ZF equalizer is utilized to remove the ISI. Finally, the transmit symbol is decoded with the decision metric.

\begin{figure}
\centering
\includegraphics[width=80mm]{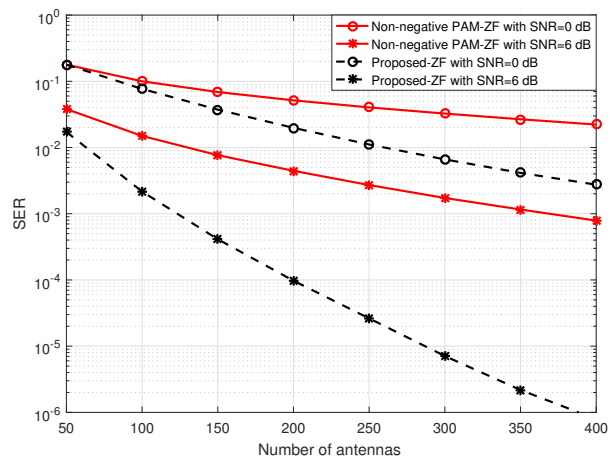}
\caption{SER versus the number of receive antennas at various SNRs with $K=4$.}
\label{Number_of_antennas}
\end{figure}

Fig. \ref{Number_of_antennas} plots SERs with different numbers of receive antennas at $\text{SNR}=\text{0~dB}$ and $\text{6~dB}$, where a constellation size of $K=4$ is employed. First of all, a remarkable performance gap between $P_{e\_opt}$ and $P_{e\_pam}$ exhibits the benefit of our proposed constellation design. As expected, the logarithmic SER decreases almost linearly with $M$ for both constellations. This observation proves the effectiveness of the approximations in \eqref{SER_optimal_antenna} and \eqref{SER_pam_antenna}, as well as demonstrating the huge potential of massive receiver array in non-coherent SIMO systems. Besides, as the number of antennas increases, the error probability of the proposed constellation decreases at a much faster rate than the PAM, which verifies the result in \eqref{inequation1_antenna}. Furthermore, \eqref{inequation1_antenna} indicates that the slope of SER versus $M$ would become larger at higher SNRs, which is also clearly demonstrated in Fig. \ref{Number_of_antennas}. 

\begin{figure}
\centering
\includegraphics[width=80mm]{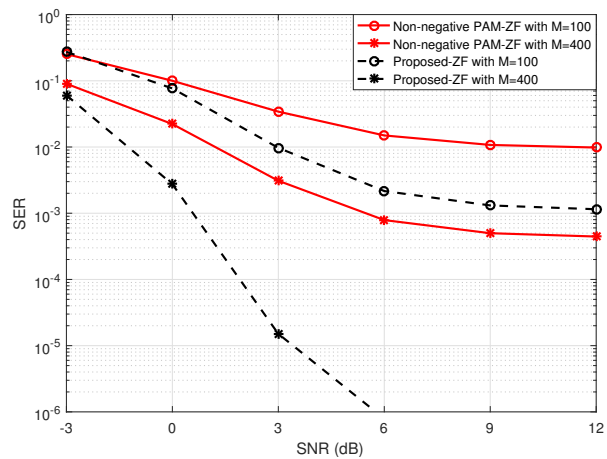}
\caption{SER versus SNR with $M=100$ and $M=400$, where $K=4$.}
\label{SNR_SER}
\end{figure}

\begin{figure}[h]
\centering
\includegraphics[width=80mm]{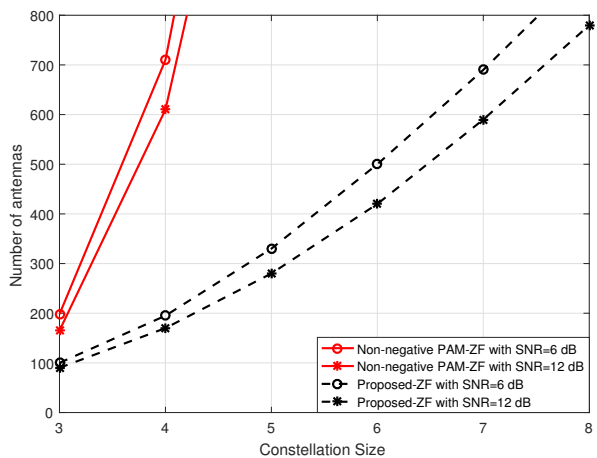}
\caption{Number of receive antennas versus the constellation size to achieve ${\rm SER} = {10^{ - 4}}$ at various SNRs.}
\label{Constellation_Size}
\end{figure}

More importantly, the required number of antennas for our design to meet a predefined SER can be decreased compared with PAM. To be more specific, Table \ref{table2} lists the number of receive antennas required in a variety of scenarios. For instance, 200 antennas is needed to achieve $\text{SER} = {10^{ - 2.5}}$ at $\text{SNR}=0~\text{dB}$ for the proposed constellation design, whereas the number of antennas for non-negative PAM to achieve the same performance is 400. Therefore, a great deal of cost in hardware implementation can be saved. This saving becomes more pronounced if the SNR increases or a lower SER is required, as Table \ref{table2} shows the required numbers of antennas in the other two settings can be reduced by 57.1\% and 66.7\%, respectively. 

Fig. \ref{SNR_SER} reports the simulated SER versus the SNR with various numbers of antennas. Obviously, the rate of descent of our constellation design is larger than that of non-negative PAM, the same as the analytical results in \eqref{inequation_noise}. Moreover, both $P_{e\_opt}$ and $P_{e\_pam}$ converge to non-zero constants as the SNR further grows. This error floor effect is caused by a finite $M$ and ISI channel. As a result, simply raising the transmit power cannot make the problem go away. Although employing more antennas can lead to a better performance, careful attention should be paid to address the balance between system performance and cost. It is also worth noting that $P_{e\_opt}$ converges to a error probability much lower than that of $P_{e\_pam}$.

\begin{table}
\centering
\caption{The Numbers of Receiving Antennas Required to Achieve Different SERs}
\label{table2}
\begin{IEEEeqnarraybox}[\IEEEeqnarraystrutmode\IEEEeqnarraystrutsizeadd{2pt}{1pt}]{v/c/v/c/v/c/v/c/v}
\IEEEeqnarrayrulerow\\
&&&\text{SER} = {10^{ - 2.5}}&&\text{SER} = {10^{ - 2}}&&\text{SER} = {10^{ - 3}}&\\
&&&\text{SNR} = 0~\text{dB}&&\text{SNR} = 6~\text{dB}&&\text{SNR} = 6~\text{dB}&\\
\IEEEeqnarraydblrulerow\\
\IEEEeqnarrayseprow[3pt]\\
&\text{Non-negative PAM}&&400&&140&&375&\IEEEeqnarraystrutsize{0pt}{0pt}\\
\IEEEeqnarrayseprow[3pt]\\
\IEEEeqnarrayrulerow\\
\IEEEeqnarrayseprow[3pt]\\
&\text{Our Constellation}&&200&&60&&125&\IEEEeqnarraystrutsize{0pt}{0pt}\\
\IEEEeqnarrayseprow[3pt]\\
\IEEEeqnarrayrulerow\\
\IEEEeqnarrayseprow[3pt]\\
&\text{Reduced by}&&50.0\%&&57.1\%&&66.7\%&\IEEEeqnarraystrutsize{0pt}{0pt}\\
\IEEEeqnarrayseprow[3pt]\\
\IEEEeqnarrayrulerow
\end{IEEEeqnarraybox}
\end{table}

Fig. \ref{Constellation_Size} plots numbers of receive antennas needed to achieve $\text{SER} = {10^{ - 4}}$ for these two constellations. In general, a greater constellation size results in a larger data rate but less reliable transmission. As can be observed from this figure, the proposed constellation design performs significantly better than non-negative PAM. For example, with a constellation of size $K = 4$, our design needs approximately one third of the number of antennas to achieve the same SER performance compared with PAM. On the other hand, both constellation designs require more antennas to maintain $\text{SER} = {10^{-4}}$ if the constellation size increases, but the proposed one needs far less antennas than non-negative PAM.

\begin{figure}
\centering
\includegraphics[width=80mm]{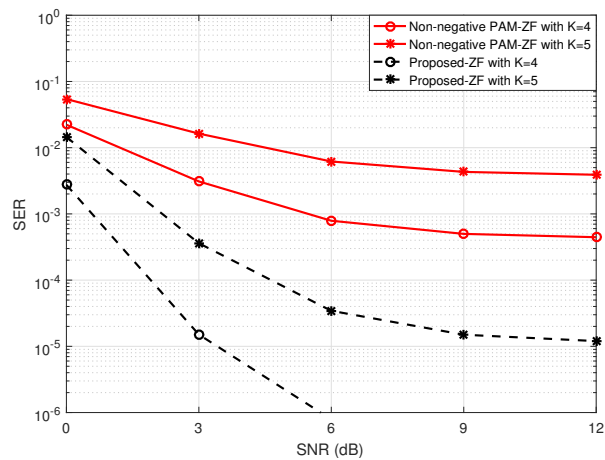}
\caption{SER versus SNR with $M=\text{400}$ for various constellations.}
\label{different_constellation_size}
\end{figure}

Fig. \ref{different_constellation_size} draws the simulated SER versus SNR for different constellations. Requiring the same number of antennas $M=400$, the proposed constellation design with a constellation size $K=5$ performs even better than a non-negative PAM constellation with $K=4$, although a higher constellation size means less reliable transmission. This observation again confirms  the huge potential of our proposed constellation design to improve the error performance.

\section{Conclusion}
This paper proposed a constellation design for a non-coherent massive SIMO system using the ED receiver in ISI channels, aiming at minimizing the symbol error probability. 

More specifically, a closed-form expression of the SER is derived, given a finite number of receiver antennas, and known channel and noise statistics, based on which an optimization problem relating to the constellation design was formulated and solved. To provide a deeper insight, we further compared the proposed constellation design with non-negative PAM in different aspects. The performance analysis indicates that the logarithms of $P_{e\_opt}$ and $P_{e\_pam}$ both linearly decrease with the number of receive antennas $M$. However, the proposed constellation design causes a much faster decline of SER versus $M$. On the other hand, the SNR analysis demonstrates that the proposed design is able to offer  the same error probability with less transmit power or receive antennas than non-negative PAM. This is especially important from the perspective of saving energy and implementation costs. Finally, an error floor for $P_{e\_opt}$ will appear at high SNRs, indicating that the performance of non-coherent massive SIMO systems is interference-limited rather than noise-limited.


\appendices

\section{Proof of Lemma1}\label{appendix A}
Suppose $\left\{ {{X_1},{X_2}, \dots ,X_n} \right\}$ is a sequence of i.i.d. random variables with ${\mathbb E}[X_i] = \mu$ and $\text{Var}[X_i] = \sigma^2 < \infty$. Then according to the Lindeberg-L\'evy CLT\cite{CLT}, as $n$ approaches infinity, the random variables $\sqrt n \left( {\frac{1}{n}\sum\nolimits_{i = 1}^n {{X_i}}  - \mu } \right)$ converge in distribution to a normal ${\cal N}(0,\sigma^2)$, i.e.
\begin{equation}
	{\displaystyle {\sqrt {n}}\left({\frac {1}{n}}\sum _{i=1}^{n}X_{i}-\mu \right)\ {\xrightarrow {d}}\ {\cal N}\left(0,\sigma ^{2}\right).}
\end{equation}
The Lindeberg-L\'evy CLT can be transformed into
\begin{equation}\label{CLT2}
\frac{1}{n}\sum\limits_{i = 1}^n {{X_i}} \;\xrightarrow{d}\;\mu + \frac{1}{{\sqrt n }}{\cal N}\left( {0,{\sigma ^2}} \right)  = {\cal N}\left( {\mu ,\frac{{{\sigma ^2}}}{n}} \right).
\end{equation}

First, the received signal at the $m$th antenna is
\begin{equation}
{y_m}( t ) = \sum\limits_{l = 0}^{L - 1} {{h_{l,m}}s( {t - l} ) + {n_m}( t )}. 
\end{equation}
Due to the mutual independence among all $h_{l,m}$, the received signals at different antennas are mutually independent. It can be derived that ${y_m}(t)$ is a complex Gaussian variable. i.e.,
\begin{equation}
{y_m}(t) \sim {\cal CN}\left( {0,\sum\limits_{l = 0}^{L - 1} {\sigma _{{h_l}}^2{{\left| {s\left( {t - l} \right)} \right|}^2}}  + \sigma _n^2} \right).
\end{equation}
Furthermore, it is shown that $\left\| {{{\mathbf{y}}(t)}} \right\|_2^2$ is a sum of $M$ i.i.d. $\left|y_{m}(t)\right|^2_{m=1,2,...,M}$ with
\begin{equation}\label{ex and var}
\begin{aligned}
\mathbb{E}\left[ {{{\left| {{y_m}( t )} \right|}^2}} \right] &= \mathbb{E}\left[ {\Re {{\left\{ {{y_m}( t )} \right\}}^2}} \right] + \mathbb{E}\left[ {\Im {{\left\{ {{y_m}( t )} \right\}}^2}} \right] \\
&= \sum\limits_{l = 0}^{L - 1} {\sigma _{{h_l}}^2{{\left| {s( {t - l} )} \right|}^2}}  + \sigma _n^2 \\
{\text{Var}}\left[ {{{\left| {{y_m}( t )} \right|}^2}} \right] &= {\text{Var}}\left[ {\Re {{\left\{ {{y_m}( t )} \right\}}^2}} \right] + {\text{Var}}\left[ {\Im {{\left\{ {{y_m}( t )} \right\}}^2}} \right] \\
&={\left( {\sum\limits_{l = 0}^{L - 1} {\sigma _{{h_l}}^2{{\left| {s( {t - l} )} \right|}^2}}  + \sigma _n^2} \right)^2}.
\end{aligned}
\end{equation}
According to \eqref{CLT2} and \eqref{ex and var}, the result in \eqref{eq6} is proved. Hence, the proof of Lemma \ref{lemma1} is concluded.

\section{Proof of Proposition \ref{prop1}}\label{appendix B}

From Fig. \ref{constellation_picture}, the range of ${p_i}$ is decoded by ${d_{L,i}}$ and ${d_{R,i}}$. As for a single transmit symbol ${p_i}$, the error probability is
\begin{equation}
{P_e}({p_i}) = {P_r}\left\{ {{p_i} \not\subset {\vartriangle _i}} \right\}
\end{equation}
where ${\vartriangle _i} = [{p_i} + w\sigma _n^2 - {d_{L,i}},{p_i} + w\sigma _n^2 + {d_{R,i}})$. Through invoking the Gaussian approximations, the probability of correct decision for each transmit symbol can be written as
\begin{equation}\label{each probability}
P({p_i}) = \frac{1}{2}\left( {\text{erf}\left( {\frac{{{d_{L,i}}}}{{\sqrt 2 \sigma_\psi ({p_i})}}} \right) + \text{erf}\left( {\frac{{{d_{R,i}}}}{{\sqrt 2 \sigma_\psi ({p_i})}}} \right)} \right)
\end{equation}
where $\sigma ({p_i})$ denotes the variance in \eqref{variance2}. 

Based on above, the average probability of errors is given by
\begin{equation}\label{SER}
\begin{array}{l}
{P_e} = 1 - \displaystyle\frac{1}{K}\sum\limits_{i = 1}^K {P({p_i})} \\
\hspace{0.5cm}= 1 - \displaystyle\frac{1}{{2K}}\sum\limits_{i = 1}^K {\left( {\text{erf}\left( {\frac{{{d_{L,i}}}}{{\sqrt 2 \sigma_\psi ({p_i})}}} \right) + \text{erf}\left( {\frac{{{d_{R,i}}}}{{\sqrt 2 \sigma_\psi ({p_i})}}} \right)} \right)}. 
\end{array}
\end{equation}
Therefore, the proof of Proposition \ref{prop1} is concluded.

\section{Solution of the quadratic equation \eqref{quadratic equation} and Proof of Proposition \ref{prop3}}\label{appendix C}
\subsection{Solution of the quadratic equation \eqref{quadratic equation}}
 The solution of  \eqref{quadratic equation} will be proved in below. 
 
 At the first, $B$ and $C$ can be represented as
\begin{align}\label{equation1}
B(T,p_i) &= \frac{{{B_1}(T,p_i)}}{{{T^2}}} + 2{B_2},\\
C(T,p_i) &= \left( {\frac{{MB_2^2}}{{{w^2}\sigma _{{h_0}}^4}} - \frac{{B_1(T,p_i)^2}}{{2{T^2}}}} \right)
\end{align}
where ${B_1(T,p_i)} = {p_i} + \sqrt 2 T{\sigma _\psi }\left( {{p_i}} \right)$ and  ${B_2} = \frac{{{w^2}\sigma _{{h_0}}^2}}{M}\left( {\sigma _n^2 + \frac{1}{K}\sum\limits_{l = 1}^{L - 1} {\sigma _{{h_l}}^2} } \right)$.

Then, ${B(T,p_i)^2} - 4AC(T,p_i)$ will be solved as
\begin{equation}\label{equation2}
\begin{array}{l}
\begin{aligned}
\displaystyle &{B(T,p_i)^2} - 4AC(T,p_i) \\
&= \frac{{B_1(T,p_i)^2}}{{{T^4}}} + \frac{{4{B_1(T,p_i)}{B_2}}}{{{T^2}}} + 4B_2^2 \\
&\quad- 4\left( {\frac{{{w^2}\sigma _{{h_0}}^4}}{M} - \frac{1}{{2{T^2}}}} \right)\left( {\frac{{MB_2^2}}{{{w^2}\sigma _{{h_0}}^4}} - \frac{{B_1(T,p_i)^2}}{{2{T^2}}}} \right)\\
\displaystyle&= \frac{2}{{{T^2}}}\left( {2{B_1(T,p_i)}{B_2} + \frac{{{w^2}\sigma _{{h_0}}^4B_1(T,p_i)^2}}{M} + \frac{{MB_2^2}}{{{w^2}\sigma _{{h_0}}^4}}} \right)\\
\displaystyle&= \frac{2}{{{T^2}}}{\left( {\frac{{\sqrt M {B_2}}}{{w\sigma _{{h_0}}^2}} + \frac{{w\sigma _{{h_0}}^2{B_1(T,p_i)}}}{{\sqrt M }}} \right)^2}
\end{aligned}	
\end{array}
\end{equation}

Based on \eqref{equation1} and \eqref{equation2}, the 
\begin{equation}\label{equation3}
\begin{array}{l}
\begin{aligned}
\displaystyle &{p_{i + 1}} = \frac{{ - B(T,p_i) \pm \sqrt {{B(T,p_i)^2} - 4AC(T,p_i)} }}{{2A}}\\
&= \frac{\displaystyle{\left( {\frac{{{B_1(T,p_i)}}}{{{T^2}}} + 2{B_2}} \right) \pm \displaystyle\frac{{\sqrt 2 }}{T}\left( {\frac{{\sqrt M {B_2}}}{{w\sigma _{{h_0}}^2}} + \frac{{w\sigma _{{h_0}}^2{B_1(T,p_i)}}}{{\sqrt M }}} \right)}}{\displaystyle{\frac{1}{{{T^2}}} - \frac{{2{w^2}\sigma _{{h_0}}^4}}{M}}}\\
&={p_i}\,\, or\,\, \frac{{\sqrt 2 T\left( {\sqrt M {\sigma _\psi }\left( {{p_i}} \right) + w\sigma _n^2 + \displaystyle\frac{w}{K}\sum\limits_{l = 1}^{L - 1} {\sigma _{{h_l}}^2} } \right) + {p_i}}}{{\sqrt M  - \sqrt 2 Tw\sigma _{{h_0}}^2}}
\end{aligned}	
\end{array}
\end{equation}
From \eqref{equation3}, through control the range of $T$, $0 < T < \sqrt {\frac{M}{2}} \frac{1}{{w\sigma _{{h_0}}^2}}$, obviously, both the solution of quadratic \eqref{quadratic equation} are real positive.

From above equation, the solution all is positive,
The process of choose the solution can be explained as that, $p_i$ could not be chosen because if it will cause that the all constellation point are equal. The other solution is our choice, thus $p_{i+1}$ can be represent as \eqref{quadratic formula}.

\subsection{Proof of Proposition \ref{prop3}}
Then it can be represent as $p_{i+1}$ is function of $T$
\begin{equation}\label{equation9}
\begin{array}{l}
\begin{aligned}
{p_{i + 1}} &= \frac{{\sqrt 2 T\left( {\sqrt M {\sigma _\psi }\left( {{p_i}} \right) + w\sigma _n^2 + \frac{w}{K}\sum\limits_{l = 1}^{L - 1} {\sigma _{{h_l}}^2} } \right) + {p_i}}}{{\sqrt M  - \sqrt 2 Tw\sigma _{{h_0}}^2}}\\
&= a\left( T \right){p_i} + b\left( T \right)
\end{aligned}	
\end{array}
\end{equation}
where 
\begin{equation*}
\begin{array}{l}
\displaystyle a\left( T \right) = \frac{{\sqrt 2 Tw\sigma _{{h_0}}^2 + 1}}{{\sqrt M  - \sqrt 2 Tw\sigma _{{h_0}}^2}}\\
\displaystyle b\left( T \right) = \frac{2{\sqrt 2 Tw}}{{\sqrt M  - \sqrt 2 Tw\sigma _{{h_0}}^2}}\left( {\sigma _n^2 + \frac{1}{K}\sum\limits_{l = 1}^{L - 1} {\sigma _{{h_l}}^2} } \right).
\end{array}
\end{equation*}
In this paper, the 
\begin{equation}
\sum\limits_{i = 1}^K {{p_i}}  = \sum\limits_{i = 1}^K {a{{\left( T \right)}^{i-1}}{p_1}}  + \sum\limits_{i = 1}^K {\left( {K - i} \right)b\left( T \right)a{{\left( T \right)}^{i - 1}}} 
\end{equation}

In order to solve this function increase or decrease, it can be divided into two part, ${a{{\left( T \right)}^i}}$ and ${b\left( T \right)a{{\left( T \right)}^{i - 1}}}$, to solve.

First, for the ${a{{\left( T \right)}^i}}$ derivative of $T$,

\begin{equation}\label{equation11}
\begin{array}{l}
\frac{{\partial a{{\left( T \right)}^i}}}{{\partial T}} = \frac{{i\sqrt 2 w\sigma _{{h_0}}^2{{\left( {\sqrt 2 Tw\sigma _{{h_0}}^2 + 1} \right)}^{i - 1}}}}{{{{\left( {\sqrt M  - \sqrt 2 Tw\sigma _{{h_0}}^2} \right)}^i}}}+ \frac{{i\sqrt 2 w\sigma _{{h_0}}^2{{\left( {\sqrt 2 Tw\sigma _{{h_0}}^2 + 1} \right)}^i}}}{{{{\left( {\sqrt M  - \sqrt 2 Tw\sigma _{{h_0}}^2} \right)}^{i + 1}}}}
\end{array}
\end{equation}

Second, for the ${b\left( T \right)a{{\left( T \right)}^{i - 1}}}$ derivative of $T$ 
\begin{equation}\label{equation12}
\begin{array}{l}
\frac{{\partial b\left( T \right)a{{\left( T \right)}^{i - 1}}}}{{\partial T}} = \left( {\sigma _n^2 + \frac{1}{K}\sum\limits_{l = 1}^{L - 1} {\sigma _{{h_l}}^2} } \right)\frac{{4{w^2}\sigma _{{h_0}}^2T{{\left( {\sqrt 2 Tw\sigma _{{h_0}}^2 + 1} \right)}^{i - 1}}}}{{{{\left( {\sqrt M  - \sqrt 2 Tw\sigma _{{h_0}}^2} \right)}^{i + 1}}}}\\
+ 2\sqrt 2 w\left( {\sigma _n^2 + \frac{1}{K}\sum\limits_{l = 1}^{L - 1} {\sigma _{{h_l}}^2} } \right)\frac{{{{\left( {\sqrt 2 Tw\sigma _{{h_0}}^2 + 1} \right)}^{i - 1}}}}{{{{\left( {\sqrt M  - \sqrt 2 Tw\sigma _{{h_0}}^2} \right)}^i}}}\\
+ 2\sqrt 2 w\left( {\sigma _n^2 + \frac{1}{K}\sum\limits_{l = 1}^{L - 1} {\sigma _{{h_l}}^2} } \right)\frac{{\left( {i - 1} \right)\sqrt 2 Tw\sigma _{{h_0}}^2{{\left( {\sqrt 2 Tw\sigma _{{h_0}}^2 + 1} \right)}^{i - 2}}}}{{{{\left( {\sqrt M  - \sqrt 2 Tw\sigma _{{h_0}}^2} \right)}^i}}}
\end{array}
\end{equation}

From \eqref{equation11} and \eqref{equation12},for $0 < T < \sqrt {\frac{M}{2}} \frac{1}{{w\sigma _{{h_0}}^2}}$, $\frac{{\partial a{{\left( T \right)}^i}}}{{\partial T}}$ and $\frac{{\partial b\left( T \right)a{{\left( T \right)}^{i - 1}}}}{{\partial T}}$ are larger than zero, it means that ${a{{\left( T \right)}^i}}$ and ${b\left( T \right)a{{\left( T \right)}^{i - 1}}}$ are monotone increasing to $T$, thus  $\sum\limits_{i = 1}^K {{p_i}}$ is monotone increasing with $T$.

\section{Proof of ${{{d_{L\_pam,2}^2}}}/{{{\zeta({p_{2\_pam}})} }}$ is Minimum of ${{{d_{i,pam}^2}}}/{{{\zeta({p_{i\_pam}})} }}$}\label{Appendix D}
At first, ${{{d_{i,pam}}}}/{{\sqrt {\zeta ({p_{i\_pam}})} }}$ can be divided into two parts, namely ${{{d_{R\_pam,i}}}}/{{\sqrt {\zeta ({p_{i\_pam}})} }}$ and $/{{{d_{L\_pam,i}}}}/{{\sqrt {\zeta ({p_{i\_pam}})} }}$. As can be inferred from \eqref{SER of PAM} and \eqref{Variance_Antennas}, ${d_{R\_pam,i}} = {d_{L\_pam,i + 1}}$ and $\zeta ({p_{i\_pam}})$ increase with ${p_{i\_pam}}$. Thus, the following relationship holds
\begin{equation}\label{eq46}
\frac{{{d_{R\_pam,i}}}}{{\sqrt {\zeta ({p_{i\_pam}})} }} > \frac{{{d_{L\_pam,i + 1}}}}{{\sqrt {\zeta ({p_{i + 1\_pam}})}}}.
\end{equation}

Given \eqref{eq46}, we need to prove that ${{{d_{L\_pam,2}}}}/{{\sqrt {\zeta ({p_{2\_pam}})} }}$ is the minimum value of ${{{d_{L\_pam,i + 1}}}}/{{\sqrt {\zeta ({p_{i + 1\_pam}})} }}$, which can be expressed as a function of $i$
\begin{equation}
\alpha (i) = \displaystyle \frac{{{w^2}M}}{D}\cdot \frac{{{i^2} + i + \frac{1}{4}}}{{{i^4} + \frac{E}{{D\varepsilon }}{i^2} + \frac{F}{{D{\varepsilon ^2}}}}}{\rm{  }},~i = 1,2, \dots, K
\end{equation}
where 
\begin{align*}
D &= \displaystyle\sigma _{{h_0}}^4 \\
E &= 2\displaystyle \sigma _{{{h_0}}}^2\sigma _{n}^2 + \frac{2}{K}\sigma _{{{h_0}}}^2\sum\limits_{l = 1}^{L - 1} {\sigma _{{h_l}}^2}\\  
F &= \displaystyle\frac{1}{{{K^2}}}\sum\limits_{l = 1}^{L - 1} {\sigma _{{h_l}}^4}  + \frac{1}{{{K^2}}}\sum\limits_{l \ne l' \ne 0} {\sigma _{{h_l}}^2} \sigma _{{h_{l'}}}^2 + \frac{2}{K}\sum\limits_{l = 1}^{L - 1} {\sigma _{{h_l}}^2\sigma _n^2 + \sigma _n^4} \\
\varepsilon  &=\displaystyle \frac{6}{{(K - 1)(2K - 1)}}. 
\end{align*}
To find the minimum value, we take the first derivative of $\alpha (i)$ with respect to $i$       
\begin{equation}\label{derivative of a(i)}
\frac{{\partial \alpha (i)}}{{\partial i}} = \displaystyle\frac{{{w^2}M}}{D}\cdot\frac{{g(i) - f(i)}}{{{{\left( {{i^4} + \frac{E}{{D\varepsilon }}{i^2} + \frac{F}{{D{\varepsilon ^2}}}} \right)}^2}}}{\rm{   }},~i = 1,2, \dots, K
\end{equation}
where $g(i) = \frac{{2F}}{{D{\varepsilon ^2}}}i + \frac{F}{{D{\varepsilon ^2}}}$, $f(i) = 2{i^5} + 3{i^4} + {i^3} + \frac{E}{{D\varepsilon }}{i^2} + \frac{E}{{2D\varepsilon }}i$.

Equation \eqref{derivative of a(i)} indicates that both $g(i)$ and $f(i)$ are monotonically increasing functions. There are three cases concerning the monotonicity of function $\alpha (i)$, namely, increase first and then decrease, monotonically increase, and monotonically decrease. In either case, one can find the minimum value by comparing the values of endpoints. As a result, the ratio of endpoints $\alpha (1)$ and $\alpha (K)$ is
\begin{equation}\label{K and 1}
\frac{{\alpha (K)}}{{\alpha (1)}} = {\left( {\frac{{2K + 1}}{3}} \right)^2}G(K)
\end{equation}
where $G(K) = \frac{{F{{(K - 1)}^2}{{(2K - 1)}^2} + 6E(K - 1)(2K - 1) + 36D}}{{36D{K^4} + F{{(K - 1)}^2}{{(2K - 1)}^2} + 6E{K^2}(K - 1)(2K - 1)}}$.

It is evident from \eqref{K and 1} that $\alpha (K)$ is greater than  $\alpha (1)$. Therefore, ${{{d_{L\_pam,i}}}}/{{\sqrt {\zeta({p_{i\_pam}})}}}$ is an increasing function or increases initially and then decreases. Therefore, $\alpha (1)$ is the minimum value and we have proved that ${{{d_{L\_pam,2}^2}}}/{{{\zeta({p_{2\_pam}})} }}$ is the minimum value of ${{{d_{i,pam}^2}}}/{{{\zeta({p_{i\_pam}})}}}$. 


\ifCLASSOPTIONcaptionsoff
  \newpage
\fi



\bibliographystyle{IEEEtran}
\bibliography{IEEEabrv,reference.bib}

\end{document}